\documentclass[12pt,a4paper]{article}

\usepackage[a4paper,margin=2.5cm]{geometry}
\usepackage[T1]{fontenc}
\usepackage{amsmath,amssymb}
\usepackage{array}
\usepackage{booktabs}
\usepackage{calc}
\usepackage{graphicx}
\usepackage{caption}
\usepackage{longtable}
\usepackage{siunitx}
\usepackage{tikz}
\usepackage[numbers,sort&compress]{natbib}
\usepackage[colorlinks=true,linkcolor=black,citecolor=black,urlcolor=blue]{hyperref}
\usetikzlibrary{arrows.meta,calc,positioning}

\graphicspath{{figures/}}
\providecommand{\tightlist}{%
  \setlength{\itemsep}{0pt}\setlength{\parskip}{0pt}}

\title{Permutationally Invariant Polynomial Descriptors as Colored Multigraphs}
\author{
  Yiping Hao$^{1,2}$, Bina Fu$^{1,2,3,*}$, Dong H. Zhang$^{1,2,3}$\\[0.7em]
  \small $^{1}$State Key Laboratory of Chemical Reaction Dynamics,\\[-0.2em]
  \small Dalian Institute of Chemical Physics, Chinese Academy of Sciences,\\[-0.2em]
  \small Dalian 116023, People's Republic of China\\
  \small $^{2}$School of Chemical Sciences, University of Chinese Academy of Sciences,\\[-0.2em]
  \small Beijing 100049, China\\
  \small $^{3}$Hefei National Laboratory, Hefei 230088, China\\[0.4em]
  \small $^{*}$Corresponding author: \href{mailto:bina@dicp.ac.cn}{bina@dicp.ac.cn}
}
\date{}

\begin{document}

\maketitle

\begin{abstract}
The combinatorial explosion in monomial counts limits the extension of global permutationally invariant polynomial (PIP) potential energy surfaces to larger molecular systems. We use graph-theoretic language to reformulate PIPs as colored multigraphs and develop a series of theoretical and computational tools.
First, we introduce an offline PIP generation algorithm whose cost has an upper bound independent of atom count for a fixed degree cutoff and number of element types.
Second, we introduce a new online PIP evaluation algorithm, termed tensor-network descent.
For the full PIP set through degree \(d\) in a system of \(M\) atoms, the evaluation complexity for \(1\le d\le14\) is \(O\!\left(f(d)M^{\left\lfloor(\sqrt{1+8d}+1)/2\right\rfloor}\right)\).
At the commonly used cutoff of \(d=5\), the online evaluation complexity decreases from the \(M^{10}\) scaling of conventional explicit term-by-term computation to \(M^3\).
Numerical tests using the full PIP set through degree 5 show that our method achieves speedups of several orders of magnitude on most MD17 systems and remains computationally affordable for systems containing more than one thousand atoms.
The method has a native tensor form, supports automatic differentiation, and provides a unified theoretical construction of four classes of PIP descriptors defined consistently across systems---global, atom-centered, bond-centered, and cluster-centered---establishing a representation and computational foundation for PIP potentials for large systems and for transfer across systems.
Together, these results show that PIPs are neither inherently restricted to small systems nor intrinsically nontransferable.

\end{abstract}

\section{Introduction}\label{sec:introduction}

A potential energy surface (PES) is the ``fuel'' for molecular dynamics simulations: its energy and gradient govern nuclear motion.
Decades ago, classical methods such as polynomial expansions and interpolation already enabled the construction of highly accurate PESs for low-dimensional systems containing three or four atoms \cite{Murrell1984PotentialEnergy,Ho1996RKHS}.
As the number of atoms increases, however, the curse of dimensionality causes the costs of configuration sampling and function representation to rise rapidly, severely limiting the extension of these methods to larger systems.
The introduction of machine-learning methods, including neural networks, opened new avenues for constructing high-dimensional PESs \cite{Sumpter1992NeuralPES,Blank1995NeuralPES}.
From high-dimensional neural network potentials to more recent continuous-filter and equivariant message-passing models, machine-learned PESs have continued to develop, providing efficient tools for accurate simulations of large systems \cite{Behler2007HDNNP,Schutt2018SchNet,Batzner2022NequIP,Batatia2022MACE}.
These methods for large systems typically use local or semilocal representations, with distance cutoffs controlling the spatial extent of atomic environments and the computational cost.
At the same time, constructing a unified, highly accurate global PES for an entire molecular configuration remains an important research direction for chemical reactions, molecular dissociation, strongly coupled vibrations, and large-amplitude intramolecular rearrangements.

Permutationally invariant polynomials (PIPs) are widely used for fitting PESs in gas-phase reaction dynamics and molecular spectroscopy \cite{Braams2009PIP,Qu2018PIPReview}.
Their original construction can be understood as first expanding the energy as a polynomial in pairwise distance variables, then requiring monomials related by permutations of like atoms to share the same coefficient, thereby expressing the energy as a linear combination of monomial orbit sums; each such orbit sum, with every distinct monomial included once, is a PIP \cite{Xie2010MSA}.
Subsequent PIP-NN methods use these polynomials as neural network inputs, combining an exactly permutation-invariant representation with nonlinear regression \cite{Jiang2013PIPNN,Li2013PIPNN2}.
Global PIPs require neither a prescribed atomic-energy decomposition nor a fixed local cutoff, and offer exact symmetry, smooth differentiability, and systematic approximation.
When strictly monotonic continuous transforms of interatomic distances are used as pair variables, the Stone--Weierstrass theorem\cite{Rudin1976Principles} guarantees that PIPs of increasing degree can uniformly approximate continuous permutation-invariant potential energy functions to arbitrary accuracy on compact configuration domains.
For gas-phase reactions involving small systems, PIP-based PESs can combine very high fitting accuracy with efficient online evaluation, while their associated neural networks can require relatively few parameters \cite{Houston2022PIPGradients,Jiang2013PIPNN,Li2013PIPNN2}.

PIPs have long been regarded as lacking scalability and being limited to small molecules, primarily because of the combinatorial explosion of monomials.
For a system containing \(M\) atoms, there are \(K=\binom{M}{2}\) pair variables and \(\binom{K+d}{d}\) monomials of degree at most \(d\), corresponding to asymptotic growth of \(M^{2d}\).
Conventional offline generation of PIP expressions requires identifying the orbits induced by permutations of like atoms within this enormous set of monomials and writing each independent PIP as a sum of monomials over specific atom labels.
As the atom count and polynomial degree increase, the costs of offline generation, formula storage, and term-by-term online evaluation all grow sharply.
Meanwhile, conventional PIP formulas are built directly from the variable set and permutation group of a particular system; changing the molecular composition or atom count also changes the variable space, group action, and analytic expressions.
This makes it difficult to share fixed feature semantics across molecules as modern machine-learning potentials do, and most PIP PESs are tailored to individual systems.


Research on expression generation, numerical evaluation, and descriptor construction has extended the molecular sizes accessible to PIP methods.
For expression generation, Moberg and Jasper developed automated, parallel software for generating PIPs \cite{Moberg2021PIPUnrestricted}.
An algorithm based on permutation-group generators and graph connectivity analysis has pushed the generation of explicit fifth-degree PIP expressions to systems of approximately 30 atoms, 
close to the largest sizes reported to date \cite{Hao2025PIPFIAlgorithms}.
For online evaluation, Houston et al. used Mathematica to generate Fortran code for reverse differentiation, 
enabling efficient force evaluation and joint fitting of energies and gradients \cite{Houston2022PIPGradients}.
More recently, JaxPIP, developed by Li et al., 
organizes explicit polynomial evaluation into tensor operations suited to batched GPU computation and integrates it with modern automatic differentiation, 
with tests on systems containing 3--9 atoms \cite{Li2026JaxPIP}.
The descriptors themselves can also be modified to reduce polynomial construction and evaluation costs.
Among methods that retain a global molecular description, the fundamental invariant neural network (FI-NN) approach introduced by our group replaces the complete linear PIP basis with algebraic generators \cite{Shao2016FINN,Chen2020FINN2,Fu2023FIReview}.
Recursive separating invariant neural networks (RSI-NNs) go further by adopting orbit separation as the design objective and constructing descriptors from low-correlation templates and recursive radial bases \cite{hao2026SRI}.
These methods have been used to fit global PESs for systems containing approximately 20 atoms.
Fragment-based representations restrict polynomial expansions to smaller components, 
allowing larger molecules to be treated \cite{Conte2020EfficientPIP,Qu2024C14H30}.
Li et al. extended MB-PIPNet to covalently bonded molecules, using PIPs to describe fragment internal structures and their environments and neural networks to predict fragment energy contributions.
Its accuracy and computational efficiency were demonstrated for n-tetradecane, a 44-atom molecule \cite{Li_YuQiAlkanes}.
Although these studies have substantially extended the applicability of PIPs and their variants, explicit monomial expansions of complete global PIP bases still face combinatorial growth.
For global methods without fragmentation or local cutoffs, the practical upper limit on system size generally remains around 20 atoms; further increases in atom count and polynomial degree rapidly raise generation, storage, and evaluation costs.

Here we identify an exact one-to-one correspondence between colored multigraphs (CMGs) and PIPs: each atom index occurring in a monomial corresponds to a vertex, vertex colors record chemical elements, and edge multiplicities record the powers of the corresponding pair variables.
After specific atom labels are removed, the graph pattern retains the elemental composition, index incidences, and power structure of the monomial, while relabeling like atoms corresponds to graph isomorphisms that preserve colors and edge multiplicities.
Thus, subject to elemental capacity constraints, PIP monomial orbits are in one-to-one correspondence with colored-multigraph isomorphism classes; the same principle has previously been used in the study of graph-invariant algebras to investigate graphs through algebraic invariant theory \cite{Thiery2000GraphInvariants}.
We use this correspondence in the reverse direction to unify PIP generation, numerical evaluation, and cross-system description within a single graph representation.
Reformulating PIPs as colored multigraphs allows a range of existing theoretical tools from graph algorithms to be introduced, with minor adaptations, for systematic use in PIP generation, algebraic transformations, and numerical evaluation.

First, we address PIP formula generation.
Through the correspondence above, generating a PIP basis becomes a problem of enumerating colored-multigraph isomorphism classes, with successive edge addition, graph-isomorphism deduplication, and element coloring directly yielding all graph patterns corresponding to PIPs at each degree.
A degree-\(d\) monomial involves at most \(2d\) atoms, so for fixed element types and a fixed degree cutoff, the number of graph patterns has an upper bound independent of the system's atom count; a specific system need only exclude patterns for which its elemental capacities are insufficient.
Once each element has sufficiently many atoms, every pattern can be realized, further increases in atom count produce no new PIP patterns, and the computational resources required for offline generation reach a constant plateau.
We perform generation at the graph-pattern level without expanding each pattern into a sum of monomials over specific atom labels, thereby avoiding the combinatorial growth of explicit formula storage.
Single-threaded numerical tests show that, after saturation, generating the complete set of graph patterns through fifth order for water systems consistently takes only about \(3\,\mathrm{s}\); even when the specified system capacity exceeds three million atoms, both the pattern count and generation time remain essentially unchanged.

Second, we address the online evaluation of PIP formulas.
We transform the labeling sum of a graph pattern into a dual tensor-network problem: first relaxing the requirement that indices for like atoms be distinct gives a graph-homomorphism sum that can be evaluated directly by tensor contraction, and M\"obius inversion on the partition lattice of same-color vertices then recovers the original PIP exactly through quotient-graph descent \cite{Lovasz2012GraphLimits,Curticapean2017Homomorphisms}.
Variable elimination reorganizes the summation order for graph-homomorphism evaluation, with the computational complexity controlled by graph treewidth \cite{Dechter1999BucketElimination,MarkovShi2008TensorNetworks}; for the complete set of PIPs through degree \(d\) for \(M\) identical atoms, the leading system-size complexity in the range \(1\le d\le14\) verified here by exhaustive computation is
\[
\mathcal O\!\left(f(d)M^{\left\lfloor(\sqrt{1+8d}+1)/2\right\rfloor}\right),
\]
where \(f(d)\) is a relatively small constant independent of system size.
At the commonly used fifth-order cutoff, the leading complexity of evaluating the complete PIP set with respect to the atom count \(M\) decreases from the \(M^{10}\) scaling of conventional explicit term-by-term evaluation to \(M^3\).
Numerical tests on MD17 systems yield online evaluation speedups of approximately \(14.6\) to \(6.6\times10^4\), extending the practical evaluation of the complete global PIP set through degree five from just over twenty atoms with conventional explicit expansions to systems containing more than one thousand atoms.
This tensorized computation can also be integrated directly into modern machine-learning frameworks, with reverse-mode automatic differentiation providing gradients with respect to the pairwise input variables \cite{Baydin2018AD}.

Finally, we consider cross-system PIP modeling.
The same colored multigraph can be labeled in systems with different atom counts and elemental compositions: changes in the system affect only the index summation domains, while the meaning of the graph pattern remains unchanged.
Thus, for fixed element types and truncation rules, different systems can naturally share a set of PIP descriptors with fixed feature semantics.
Building on this, we use black and white coloring to distinguish retained center indices from summed environment indices, constructing four classes of PIP descriptors defined consistently across systems: global molecular, atom-centered, bond-centered, and cluster-centered descriptors, which provide a unified representation and computational foundation for transferable PIP potentials.
We give theoretical constructions of these descriptors and numerically validate global PIP generation and evaluation; cross-system potential learning is left for future work.
Notably, existing atomic PIPs \cite{Allen2021aPIP} and transferable many-body PIP potentials for alkanes \cite{Qu2025TransferableAlkanes} both use symmetrization within small atom clusters followed by summation over atomic combinations.
For the polynomial parts expressed in the same pair variables, this construction is mathematically equivalent to the two-stage labeling sum in our graph-theoretic formulation.
These studies have already achieved cross-system sharing of features and parameters in specific systems and verified the resulting transferability.
We further provide methods for graph-pattern generation, complexity analysis, and tensor-contraction evaluation, establishing a computational basis for extending such approaches to larger and more general systems.

Our results show that retaining a complete PIP basis need not incur the combinatorial cost of explicit monomial expansion; the colored-multigraph representation unifies PIP generation, evaluation, and cross-system description, opening new avenues for constructing PIP potentials for large systems and for transfer across systems.

\section{Methods}\label{sec:methods}

\subsection{1 PIPs represented by colored multigraphs}\label{sec:cmg-pip}

\subsubsection{1.1 Molecular systems, pair variables, and permutation groups}\label{sec:molecular-system}

Consider a molecular system containing \(E\) chemical elements. Let \(M_a\) denote the number of atoms of element \(a\), and let the set of all atom labels be

\[
\mathcal A=\bigsqcup_{a=1}^{E}\mathcal A_a,
\qquad |\mathcal A_a|=M_a .
\]

Only atoms of the same element may be permuted, so the allowed permutation group is

\[
G=\prod_{a=1}^{E}S_{M_a}.
\]

For every pair of distinct atoms \(i,j\), define a pair variable \(x_{ij}=x_{ji}\). It may be the distance \(r_{ij}\), a Morse-type variable, an inverse distance, or another radial function that depends only on the atom pair. After blocking the variables by element type, all pair variables can be organized as a collection of second-order tensors, or matrices,

\[
X^{(a,b)}\in\mathbb R^{M_a\times M_b},
\qquad
X^{(a,b)}_{ij}=x_{ij}.
\]

For \(a=b\), \(X^{(a,a)}\) is a symmetric square matrix with diagonal elements defined by \(x_{ii}=0\). Matrices between distinct elements satisfy \(X^{(b,a)}=(X^{(a,b)})^\mathsf T\). A monomial of total degree \(d\) can be written as

\[
m(\mathbf x)=\prod_{i<j}x_{ij}^{\alpha_{ij}},
\qquad
\sum_{i<j}\alpha_{ij}=d ,
\]

where each \(\alpha_{ij}\) is a nonnegative integer. The group \(G\) acts on monomials by simultaneously replacing atom labels. The unnormalized orbit sum generated by a representative monomial \(m\) is

\[
\operatorname{Orb}(m)=\sum_{m'\in Gm}m' ,
\]

where every distinct monomial appears exactly once. In this article, a PIP refers to such a monomial orbit sum.

\subsubsection{1.2 From a monomial to a colored multigraph}\label{sec:monomial-to-cmg}

The index structure of any monomial in a PIP has a unique representation as a labeled colored multigraph.

Given a monomial \(m\), first construct a labeled colored multigraph \(\widetilde F_m\). Every atom label \(i\) that actually occurs in the monomial corresponds to a vertex labeled \(i\), whose color is determined by the element of that atom. If the factor \(x_{ij}\) occurs \(\alpha_{ij}\) times in \(m\), place \(\alpha_{ij}\) parallel edges between vertices \(i\) and \(j\). Because pair variables are defined only between distinct atoms, \(\widetilde F_m\) has no self-loops.

Next remove the atom labels from \(\widetilde F_m\), retaining only vertex colors, adjacency, and edge multiplicities. This gives the colored multigraph \(F_m=[\widetilde F_m]\). Here \(F_m\) denotes an isomorphism class that preserves colors and edge multiplicities; computations store a canonically labeled representative. We abbreviate an arbitrary colored multigraph as \(F\). Its total edge multiplicity,

\[
|E(F)|_{\mathrm{multi}}
=\sum_{\{u,v\}\subseteq V(F)}m_{uv}
=d,
\]

equals the algebraic degree of the monomial, while \(|V(F)|\) equals the number of atom indices that actually occur in it.

As an example, consider the following third-order PIP for ammonia:

\[P_F=x_{N_1H_1}^{2}x_{H_1H_2}+x_{N_1H_1}^{2}x_{H_1H_3}+x_{N_1H_2}^{2}x_{H_1H_2}+x_{N_1H_2}^{2}x_{H_2H_3}+x_{N_1H_3}^{2}x_{H_1H_3}+x_{N_1H_3}^{2}x_{H_2H_3}.\]

\begin{figure}[htbp]
\centering
\begin{tikzpicture}[
  panel/.style={inner sep=0pt,outer sep=0pt},
  flow/.style={-{Latex[length=2.4mm,width=1.9mm]},draw=black!65,line width=0.75pt},
  flowlabel/.style={align=center,text=black!65,inner sep=1pt}
]
\node[panel] (figonea) {\includegraphics[width=0.23\linewidth]{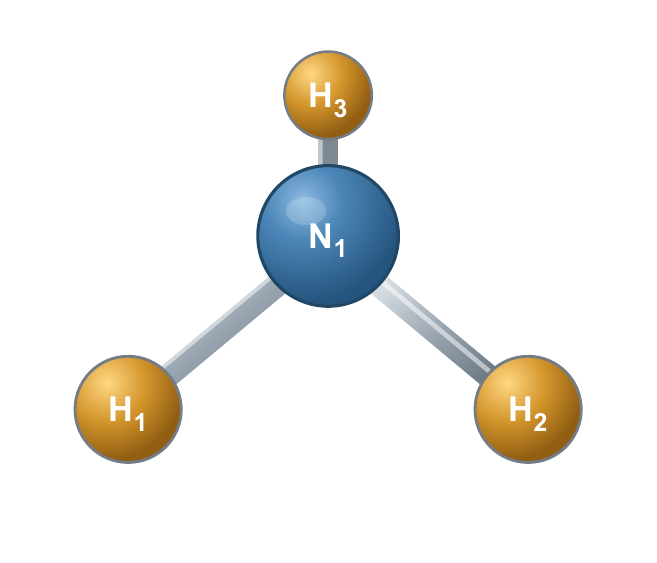}};
\node[panel,right=0.155\linewidth of figonea] (figoneb) {\includegraphics[width=0.23\linewidth]{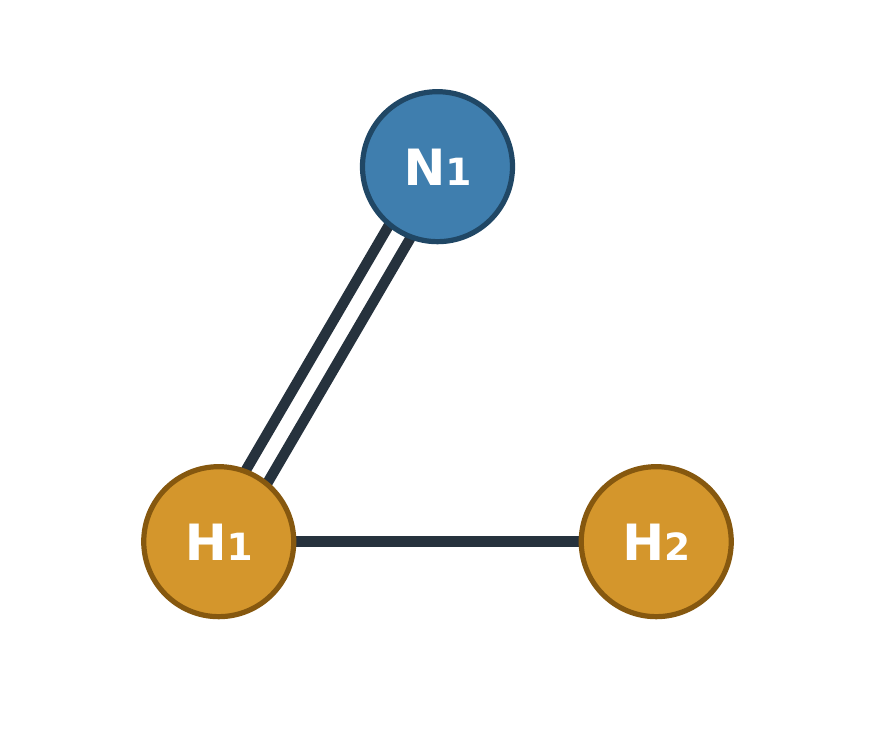}};
\node[panel,right=0.155\linewidth of figoneb] (figonec) {\includegraphics[width=0.23\linewidth]{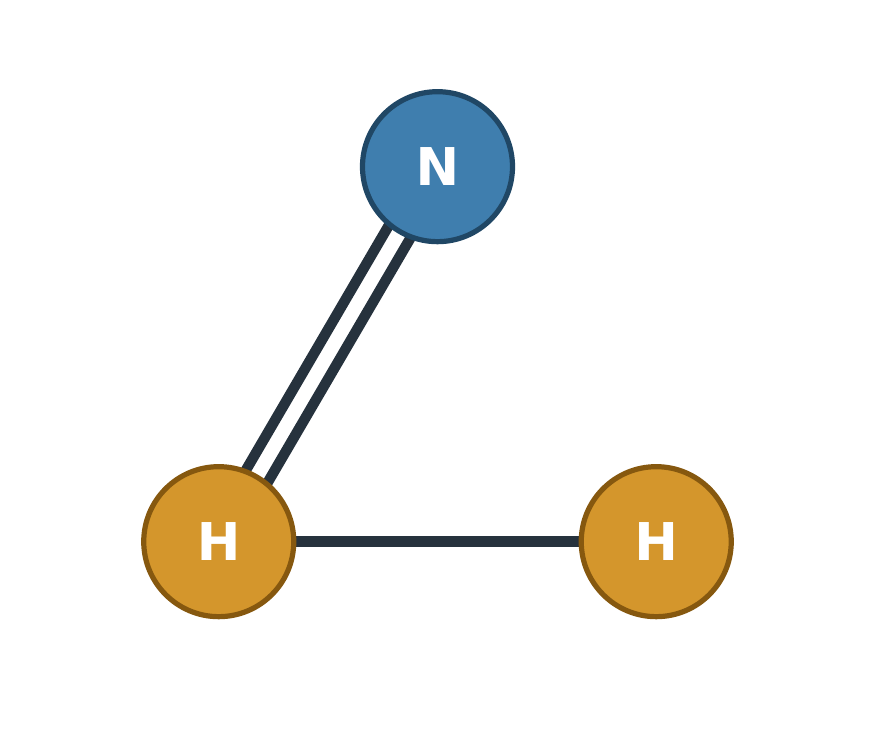}};

\node[anchor=south west,font=\bfseries\small] at (figonea.north west) {(a)};
\node[anchor=south west,font=\bfseries\small] at (figoneb.north west) {(b)};
\node[anchor=south west,font=\bfseries\small] at (figonec.north west) {(c)};

\draw[flow] (figonea.east) -- (figoneb.west);
\node[flowlabel] at ($(figonea.east)!0.5!(figoneb.west)+(0,0.62cm)$)
  {\footnotesize choose a monomial\\[-1pt]
   \(\scriptstyle m=x_{N_1H_1}^{2}x_{H_1H_2}\)};

\draw[flow] (figoneb.east) -- (figonec.west);
\node[flowlabel] at ($(figoneb.east)!0.5!(figonec.west)+(0,0.43cm)$)
  {\footnotesize forget atom labels};
\end{tikzpicture}
\caption{From a monomial of ammonia to a colored multigraph. \textbf{(a)} Ammonia with atom labels; the ball-and-stick lines indicate only the molecular geometry. \textbf{(b)} The labeled colored multigraph associated with the monomial \(x_{N_1H_1}^{2}x_{H_1H_2}\). The exponent 2 produces a double edge, and \(H_3\), which does not occur in the monomial, is absent from the graph. \textbf{(c)} The colored multigraph obtained after removing concrete atom labels. Dark graph edges denote pair variables in the monomial, not chemical bonds.}
\label{fig:monomial-to-cmg-ammonia}
\end{figure}

Choose the monomial \(m=x_{N_1H_1}^{2}x_{H_1H_2}\). Figure~\ref{fig:monomial-to-cmg-ammonia} separates the conversion into three levels. The labeled molecular entity supplies the available atom labels. After selecting a monomial, only the atom indices that occur in it are retained and edge multiplicities are determined by variable exponents. Finally, the concrete atom labels are forgotten, yielding a colored-multigraph isomorphism class that retains element colors, adjacency, and edge multiplicities.

Every graph isomorphism that preserves colors and edge multiplicities is equivalent to a relabeling of like atoms in the system. If two monomials lie in the same \(G\)-orbit, their colored multigraphs are necessarily isomorphic. Conversely, distinct PIPs correspond to nonisomorphic colored multigraphs. A nonzero PIP orbit is uniquely determined by its corresponding colored-multigraph isomorphism class without isolated vertices.

The correspondence between orbit sums of uncolored multigraphs and graph invariants can be traced to work on rings of graph invariants \cite{Thiery2000GraphInvariants}. Here vertex colors encode chemical elements, and the correspondence is used in reverse to generate and evaluate PIPs.

\subsubsection{1.3 Recovering a PIP from a graph pattern}\label{sec:recover-pip}

Every colored multigraph can be converted back into its PIP by summing over all legal labelings. For a particular molecular system, the vertices of an unlabeled colored multigraph may be relabeled only by atom labels of the same color, and labels assigned to vertices of the same color must be pairwise distinct. Every legal labeling produces one monomial. Summing the monomials over all legal labelings gives

\[
J_F(\mathbf x)=\sum_{\phi\ \text{legal labeling}}
\prod_{\substack{\{u,v\}\subseteq V(F)\\m_{uv}>0}}
x_{\phi(u)\phi(v)}^{m_{uv}} .
\]

\begin{figure}[tbp]
\centering
\includegraphics[width=0.9\linewidth]{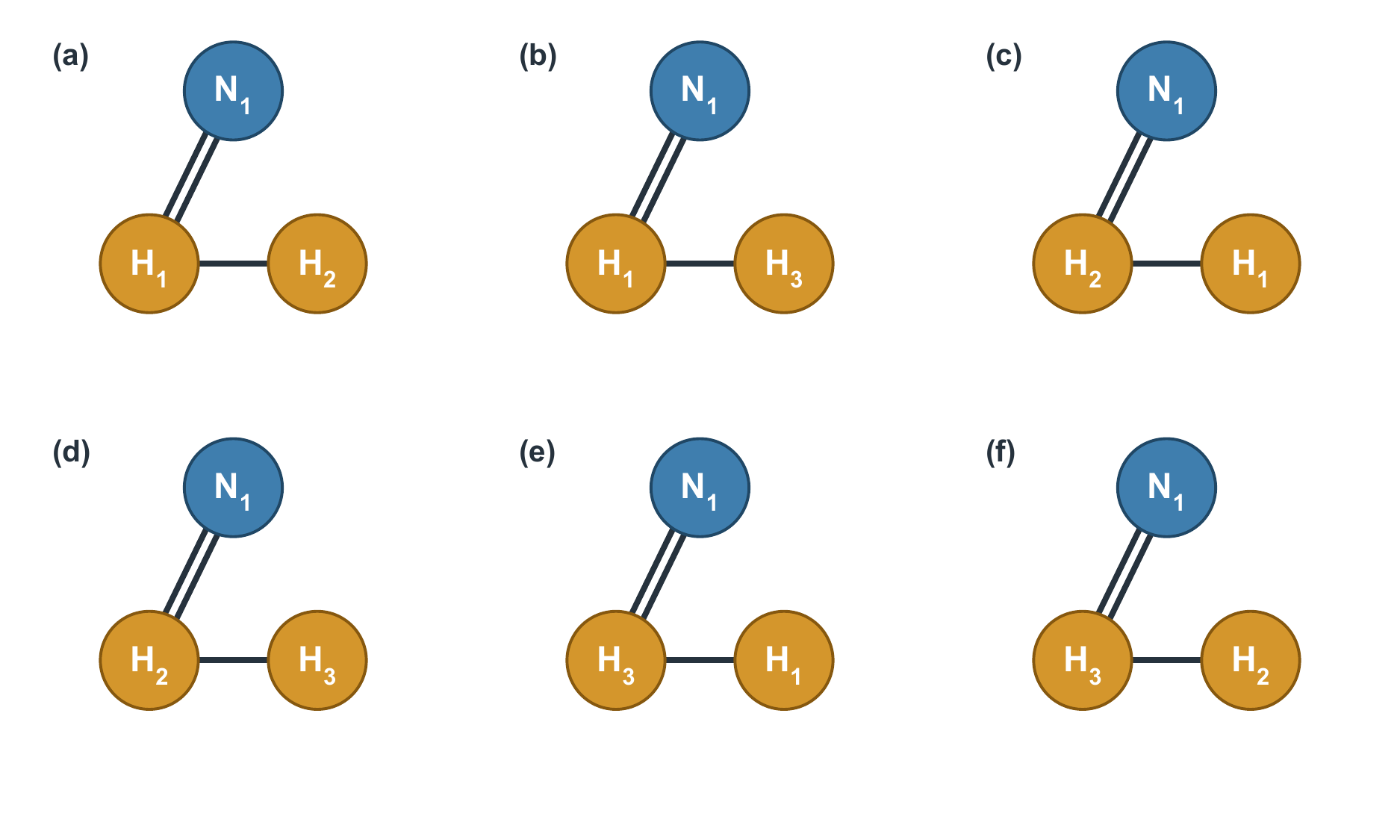}
\caption{The six legal relabelings of an asymmetric ammonia graph pattern. The nitrogen vertex is fixed as \(\mathrm{N}_1\). The two inequivalent hydrogen vertices receive an ordered pair of distinct labels from \(\mathrm{H}_1,\mathrm{H}_2,\mathrm{H}_3\), giving \(3\times2=6\) labelings and six distinct monomials.}
\label{fig:relabel-ammonia-six}
\end{figure}

Figure~\ref{fig:relabel-ammonia-six} uses an asymmetric graph pattern in ammonia to show all six legal relabelings obtained when two hydrogen vertices receive distinct labels selected from three hydrogen atoms.

\begin{figure}[tbp]
\centering
\includegraphics[width=0.9\linewidth]{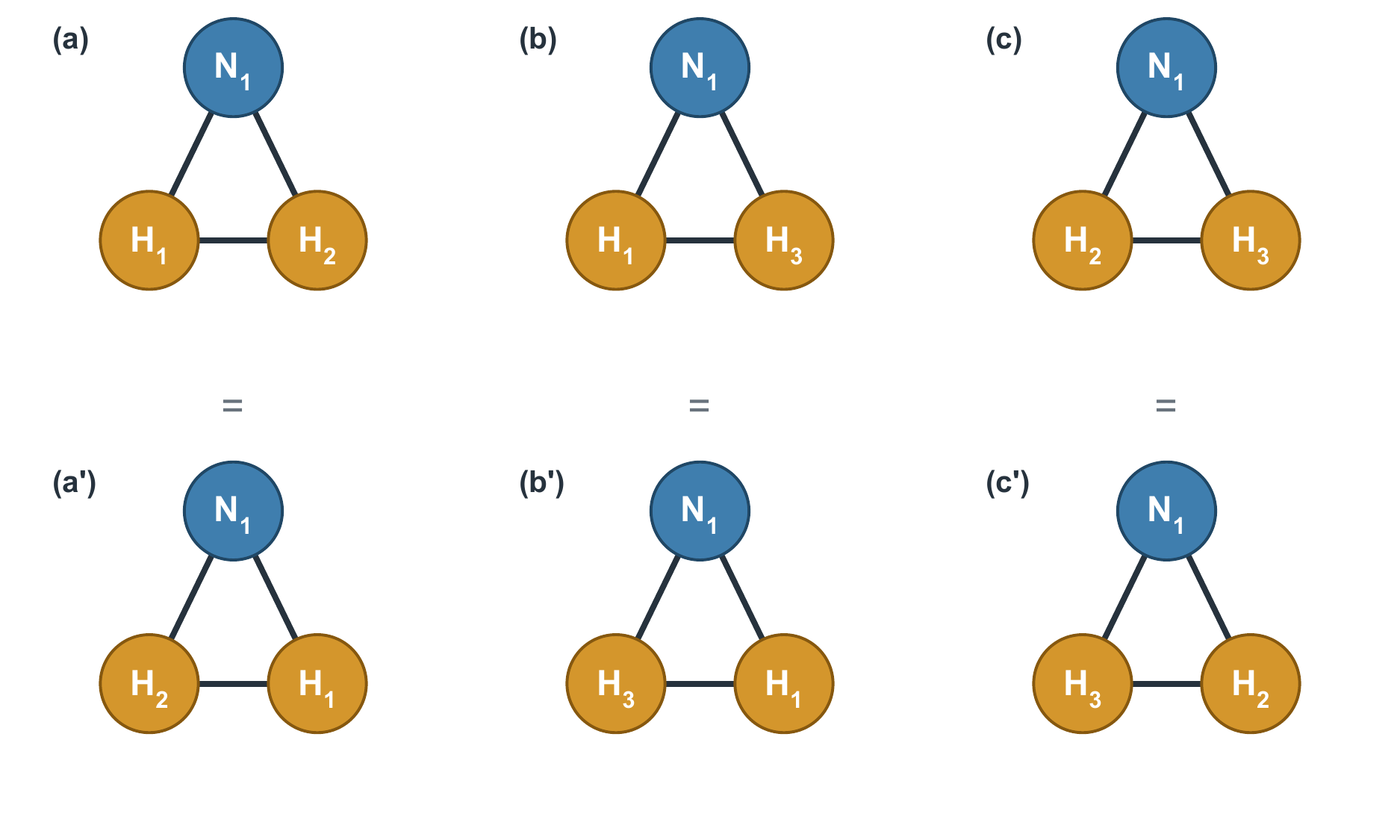}
\caption{An \(\mathrm{NHH}\) triangular graph pattern with exchange symmetry between the hydrogen vertices. In each column, the lower labeling is obtained from the upper one by exchanging the two hydrogen vertices. Because this exchange is a color- and multiplicity-preserving graph automorphism, each paired set of relabelings produces the same monomial. The six legal labelings therefore produce only three distinct monomials, consistent with \(|\operatorname{Aut}_{\chi}(F)|=2\).}
\label{fig:relabel-nhh-triangle-six}
\end{figure}

This expression is an integer multiple of the PIP rather than the PIP itself, because distinct legal labelings may yield the same monomial. For example, exchanging the endpoints of a same-color single-edge graph leaves the graph unchanged, so each monomial occurs twice. Figure~\ref{fig:relabel-nhh-triangle-six} gives a concrete example: exchanging the labels on two equivalent hydrogen vertices does not change the resulting monomial. In general, the integer multiplicity is exactly the order \(|\operatorname{Aut}_{\chi}(F)|\) of the color-preserving automorphism group of \(F\). Therefore,

\[
P_F(\mathbf x)=
\frac{J_F(\mathbf x)}{|\operatorname{Aut}_{\chi}(F)|}.
\]

The labeling enumeration can be divided into two stages. First, select \(n_a(F)\) atoms from the \(M_a\) atoms of each element; then bijectively place the selected atoms on vertices of the corresponding color. This two-stage procedure has already been used in summation based on conventional explicit expansions \cite{Qu2024C14H30,Qu2025TransferableAlkanes}. The first stage has

\[
\prod_a\binom{M_a}{n_a(F)}
\]

choices, while repetitions in the second stage are controlled exactly by \(|\operatorname{Aut}_{\chi}(F)|\).

\subsubsection{1.4 Color exhaustion and uniqueness in finite systems}\label{sec:color-exhaustion}

If, for some color \(a\),

\[
n_a(F)>M_a,
\]

then no labeling of \(F\) can satisfy both the color and distinctness requirements. In the PIP basis of a finite system, \(F\) then corresponds to no polynomial; equivalently, when retaining a unified set of all graph patterns, it may be regarded as the zero polynomial, \(P_F=0\). We call this situation color exhaustion. Once color-exhausted patterns are excluded, colored-multigraph isomorphism classes are in one-to-one correspondence with the PIPs of the system. For any colored multigraph and any molecular system, the graph necessarily defines a PIP of that system whenever color exhaustion does not occur.

\subsubsection{1.5 Support graph, interaction order, and three complexity measures}\label{sec:support-graph}

Merge parallel edges of \(F\) into single edges and temporarily ignore vertex colors. The resulting simple graph,

\[
\operatorname{supp}(F),
\]

is called the support graph, or contraction graph, of \(F\). Three quantities can be defined on it:

\begin{enumerate}
\def\labelenumi{\arabic{enumi}.}
\tightlist
\item the index count \(|V(F)|\), equal to the number of nested loops required by conventional index-by-index enumeration;
\item the interaction order, equal to the number of support edges \(|E(\operatorname{supp}F)|\); and
\item the contraction width \(\operatorname{tw}(\operatorname{supp}F)+1\), which determines the leading system-size exponent of the tensor algorithm introduced here.
\end{enumerate}

These quantities describe different aspects of evaluation complexity. The index count determines the complexity of conventional monomial enumeration and is positively correlated with interaction order, whereas contraction width controls the complexity of our new algorithm.

\subsection{2 Degree-by-degree enumeration of colored multigraphs}\label{sec:cmg-enumeration}

A PIP is determined by a monomial orbit, and monomial orbits are in one-to-one correspondence with colored-multigraph isomorphism classes. A PIP basis can therefore be enumerated directly as all colored multigraphs whose total edge multiplicity is \(d\), rather than by first generating an enormous monomial set. We first enumerate uncolored graphs degree by degree and then color them.

\subsubsection{2.1 Generating all multigraphs by edge addition}\label{sec:edge-addition}

We first ignore colors and generate, degree by degree, all isomorphism classes of multigraphs without isolated vertices. Suppose every degree-\(d\) graph has been obtained. Any degree-\((d+1)\) graph can be reduced by deleting an edge to a graph of degree at most \(d\); in reverse, it suffices to add one edge to an existing graph. There are only three types of edge addition:

\begin{enumerate}
\def\labelenumi{\arabic{enumi}.}
\tightlist
\item add an edge between two existing vertices, including increasing the multiplicity of an existing edge;
\item add an edge between an existing vertex and a new vertex; or
\item add a new isolated-edge component between two new vertices.
\end{enumerate}

After each edge addition, the candidate graph is canonicalized and deduplicated by its canonical form, yielding all uncolored degree-\((d+1)\) multigraphs. We do not seek to improve general graph-isomorphism algorithms, but use a canonicalization procedure combining WL-1 color refinement with lexicographic search. Although it is not the best available general-purpose canonical-labeling algorithm, it is simple to implement and highly efficient over the low-order graph range typical of PIPs.

\subsubsection{2.2 From uncolored graphs to element-colored graphs}\label{sec:graph-coloring}

After obtaining an uncolored multigraph \(F\), assign its vertices one of \(E\) element colors. Two colorings are considered identical only if a graph isomorphism preserves both colors and edge multiplicities. In practice, one may first enumerate color multisets, take orbit representatives of the color assignments under the automorphism group of the uncolored graph, and finally apply the color-exhaustion constraints

\[
n_a(F)\le M_a,
\qquad a=1,\ldots,E.
\]

Patterns that violate these constraints are removed.

\subsubsection{2.3 A system-size-independent upper bound on the number of patterns}\label{sec:pattern-bound}

For fixed maximum degree \(d_{\max}\) and number of colors \(E\), the number of colored-multigraph isomorphism classes before color-exhaustion filtering is bounded independently of the atom counts \(M_a\). Thus, molecules with the same element alphabet can select PIPs from the same finite pattern library regardless of their number of atoms; smaller systems merely delete more patterns through color exhaustion.

In the one-color case, the number of isomorphism classes of loopless multigraphs with no isolated vertices and exactly \(d\) edges corresponds to OEIS A050535 \cite{OEISA050535}. Its first terms are

\[
1,\ 3,\ 8,\ 23,\ 66,\ 212,\ 686,\ 2389,\ 8682,\ldots
\]

This sequence is an important reference for the degree-resolved number of PIP patterns in a single-element system. The corresponding counts of connected loopless multigraphs are given by OEIS A076864 \cite{OEISA076864}. For multiple colors, however, the total number of degree-\(d\) multigraphs grows very rapidly with the number of colors; we call this phenomenon color explosion.

\subsubsection{2.4 Obtaining PIPs from graph patterns}\label{sec:patterns-to-pips}

Section 1.3 gives the legal-labeling sum that recovers a PIP from a colored multigraph. A direct implementation admits two possibilities. One may enumerate every color-preserving injective labeling, write each labeled colored multigraph as a monomial, and combine duplicates to obtain the complete analytic expression. This simply recreates the orbit expansion and can again produce an enormous number of monomials. Alternatively, one may avoid generating an analytic expression and, during every evaluation, traverse all legal injective labelings and directly accumulate the product of edge variables. This saves formula storage but does not eliminate enumeration of label combinations or the distinct-index constraints. Neither direct approach is therefore suitable for large systems or high-order PIPs. Section 3 develops a tensor-contraction method that avoids explicit monomial expansion.

\subsection{3 Dual tensor contraction from graph-homomorphism sums to PIPs}\label{sec:dual-tensor}

\subsubsection{3.1 Graph representations of tensor networks and the dual problem}\label{sec:dual-network}

A tensor network uses a graph to represent a collection of tensors and the connectivity of their indices. In the conventional graph representation, every tensor is a vertex and its indices are edges incident to that vertex. If two tensors share an index, their vertices are connected by an edge. Summing over internal indices contracts the connected tensors.

We consider the dual situation. Every elementary tensor is second order, namely a pair-variable matrix defined in Section 1.1, while the same summation index may appear in several matrices. Each index is then represented by a vertex, each matrix factor by an edge joining two indices, and summing over a vertex eliminates the index represented by that vertex. Because the roles of tensors and indices are exchanged in the graph, we call this representation a dual tensor network. Conventional tensor networks can represent such contractions by introducing copy tensors. Conversely, a general higher-order tensor can be encoded in the dual representation by the clique formed by its index vertices, provided all edges of the clique retain their common association with that tensor object. Under this bookkeeping, the two representations are mutually convertible in expressive power. We use the dual representation of indices as vertices and second-order tensors as edges because PIPs are natively built from pair variables and the same atomic index may be shared by multiple matrix factors.

\begin{figure}[htbp]
\centering
\includegraphics[width=0.9\linewidth]{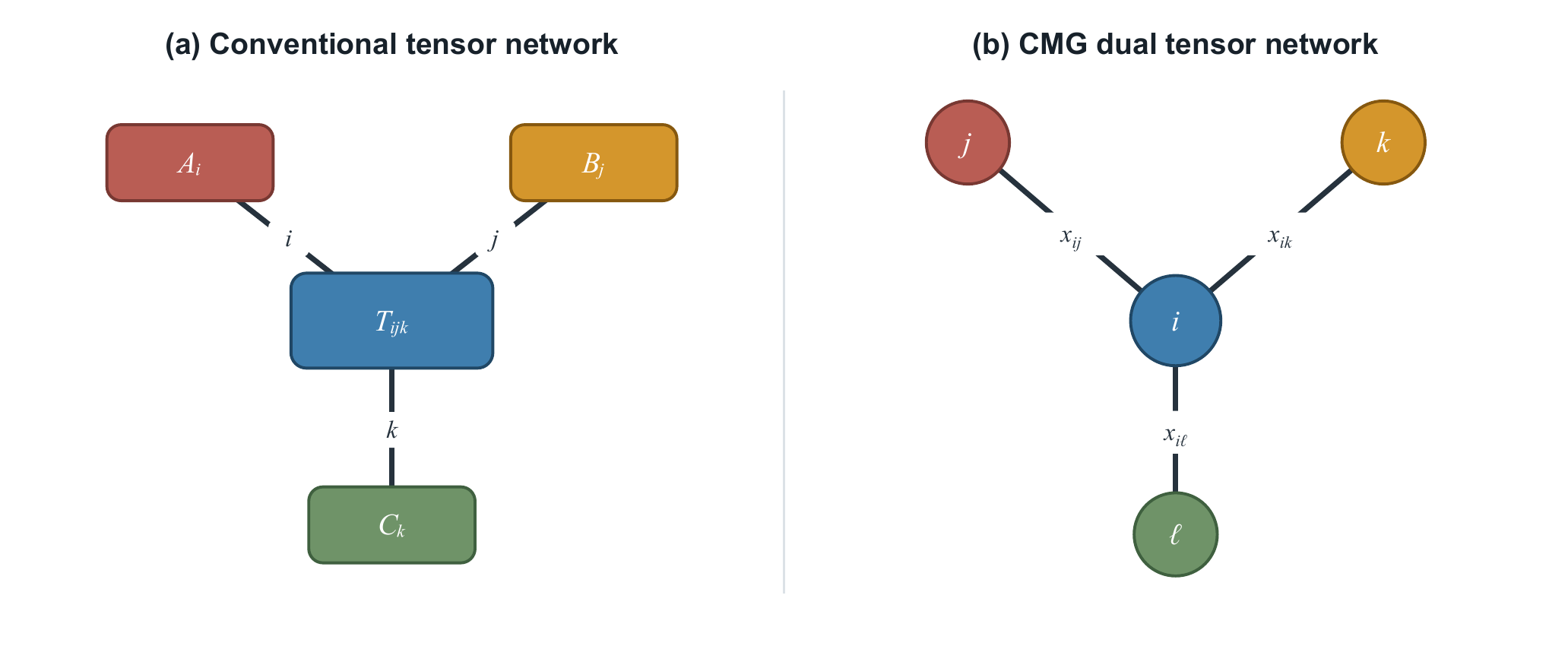}
\caption{Different incidence structures in a conventional tensor network and a CMG dual tensor network. On the left, tensors are vertices and the third-order tensor \(T_{ijk}\) carries three index edges. On the right, summation indices are colored vertices and three matrix edges share the same index \(i\). The two panels do not represent the same tensor contraction.}
\label{fig:tensor-network-dual}
\end{figure}

Figure~\ref{fig:tensor-network-dual} contrasts the two structures. In a conventional tensor network, a third-order tensor \(T_{ijk}\) can connect the three indices \(i,j,k\) simultaneously. In the dual representation, every basic tensor is a matrix, but the same index \(i\) may appear simultaneously in three or more matrix factors.

A colored multigraph can be interpreted directly as such a dual tensor network. The vertex color determines the range of the corresponding index; an edge between colors \(a\) and \(b\) represents the matrix \(X^{(a,b)}\); and an edge of multiplicity \(m\) contributes the matrix factor \((X^{(a,b)}_{ij})^m\).

\subsubsection{3.2 Weighted graph-homomorphism sums with and without distinctness constraints}\label{sec:homomorphism-sums}

The legal-labeling sum \(J_F\) from Section 1.3 requires vertices of the same color to receive pairwise distinct atom labels. This constraint cannot be expressed directly as local matrix contractions in the dual tensor network.

Remove the distinctness constraint and allow different vertices of the same color to receive the same atom label. The resulting unconstrained weighted graph-homomorphism sum is

\[
H_F(\mathbf x)
=
\sum_{\substack{\phi:V(F)\rightarrow\mathcal A\\
\phi(V_a)\subseteq\mathcal A_a}}
\prod_{\substack{\{u,v\}\subseteq V(F)\\m_{uv}>0}}
x_{\phi(u)\phi(v)}^{m_{uv}} .
\]

Under the graph representation of Section 3.1, \(H_F\) is exactly the contraction value of the dual tensor network with graph structure \(F\). Because \(x_{ii}=0\), terms in which adjacent vertices receive the same label vanish automatically, while labels of nonadjacent vertices of the same color may coincide. A disconnected graph is naturally evaluated as the product of the scalar contraction values of its connected components. Because of the extra terms generated by coincident labels, \(H_F\ne J_F\) in general. The next section explains how to recover \(J_F\) from the \(H_F\).

\subsubsection{3.3 Pattern containment and a finite partially ordered set}\label{sec:quotient-poset}

The quantity \(H_F\) is invariant under permutations of like atoms and has the same algebraic degree as \(F\). It must therefore be a linear combination of same-degree PIPs and, equivalently, of a set of \(J_Q\). The only difference between \(H_F\) and \(J_F\) is that the former permits vertices of the same color to share an atom label.

Consider a three-vertex path whose endpoints have the same color. The sum \(H_F\) contains two classes of labelings. Terms with different endpoint labels constitute \(J_F\). Terms with identical endpoint labels are equivalent to merging the endpoints into one vertex, which turns the original two edges into a pair of parallel edges. Let \(Q\) denote the resulting colored multigraph. In this example,

\[
H_F=J_F+J_Q.
\]

More generally, for any labeling in \(H_F\), vertices that receive the same atom label can be merged and edges that fall between the same pair of merged vertices become parallel edges. This produces another colored multigraph. Vertices of different colors cannot share a label. If adjacent vertices share a label, their term vanishes because \(x_{ii}=0\). Thus, only same-color nonadjacent vertices are merged in nonzero terms.

Merging preserves total edge multiplicity. If \(F\) is connected, the merged graph remains connected. Consequently, \(H_F\) for a connected degree-\(d\) graph contains only \(J_Q\) associated with other connected degree-\(d\) graphs. Collecting the coincidence patterns that produce isomorphic copies of \(Q\) gives

\[
H_F=J_F+\sum_{Q\prec F}c(F,Q)J_Q,
\]

where \(c(F,Q)\) is the number of distinct label-coincidence patterns that map \(F\) to \(Q\).

If \(Q\) can be obtained from \(F\) by a sequence of such vertex merges, we say that \(F\) entails the graph pattern \(Q\) and write \(Q\preceq F\). On connected colored multigraphs of fixed degree, this relation is a partial order:

\begin{enumerate}
\def\labelenumi{\arabic{enumi}.}
\tightlist
\item Reflexivity: merging no vertices leaves \(F\), so \(F\preceq F\).
\item Transitivity: if merging in \(F\) gives \(Q\), and merging in \(Q\) gives \(R\), the two merge operations can be combined to obtain \(R\) directly from \(F\).
\item Antisymmetry: every nontrivial merge strictly decreases the number of vertices. Hence, if both \(Q\preceq F\) and \(F\preceq Q\), no nontrivial merge can occur in either direction and \(F\) and \(Q\) are the same graph pattern.
\end{enumerate}

For fixed algebraic degree and number of colors, only finitely many graph patterns participate, so they form a finite poset.

\subsubsection{3.4 Descent on the quotient poset}\label{sec:poset-descent}

The finite poset can be represented as a directed acyclic graph. Each vertex represents a colored multigraph. If \(Q\prec F\) and no intermediate graph pattern lies between them, a cover edge is directed from \(F\) to \(Q\). Adding all edges implied by transitivity gives the full order relation. Every edge points toward a graph with fewer vertices, so directed cycles are impossible.

\begin{figure}[htbp]
\centering
\includegraphics[width=0.9\linewidth]{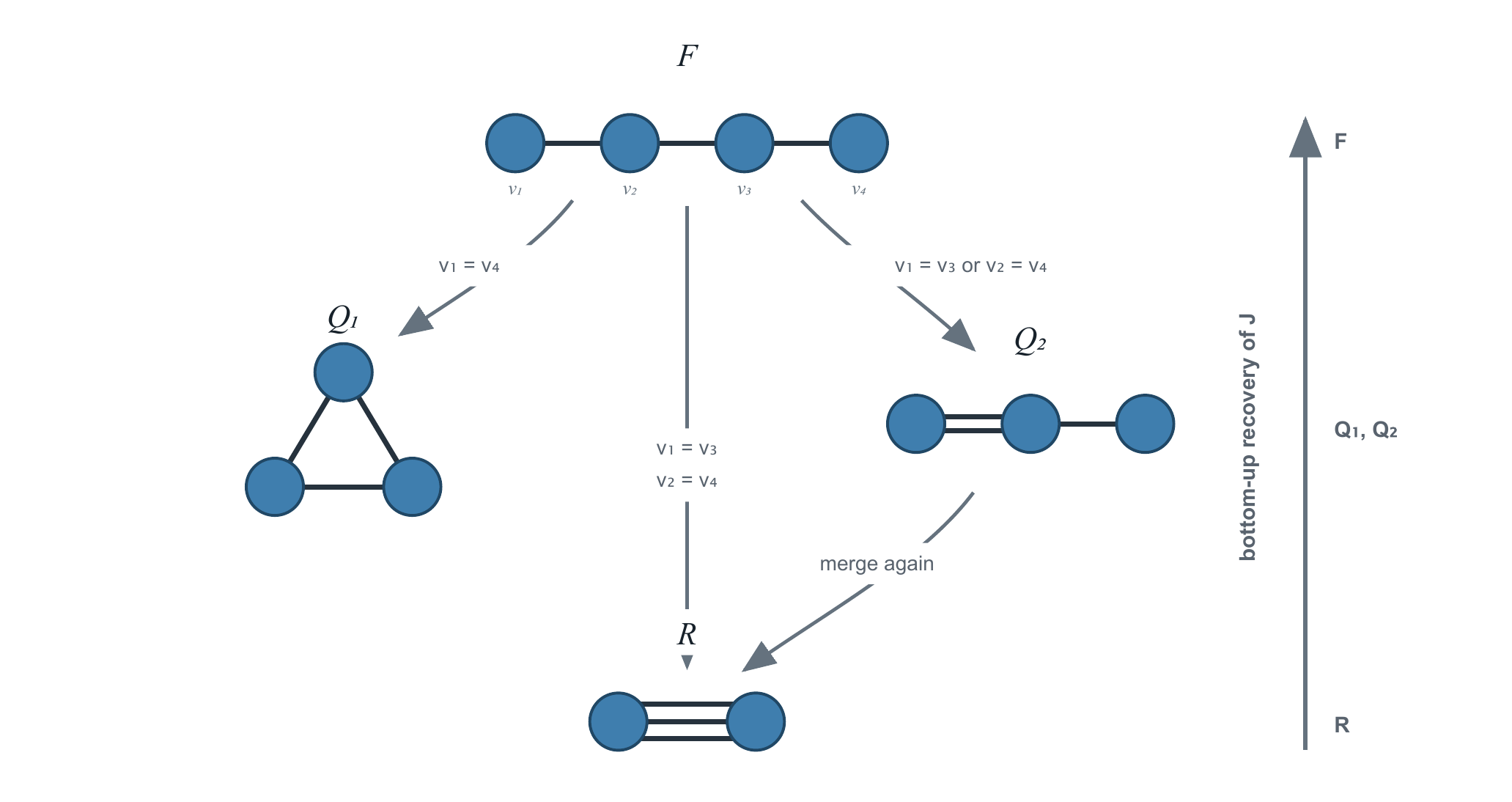}
\caption{Quotient-graph relations induced by coincident labels and the corresponding descent solution. The solid arrows on the left and right represent quotient transformations produced by one vertex coincidence. The central \(F\rightarrow R\) arrow applies \(v_1=v_3\) and \(v_2=v_4\) simultaneously. The vertical arrow at far right gives the order in which \(J\) is recovered, from \(R\) through \(Q_1,Q_2\) to \(F\). The symbols \(v_1,\ldots,v_4\) are temporary vertex names used only to explain the merge operations; they are not atom labels.}
\label{fig:quotient-poset-descent}
\end{figure}

Figure~\ref{fig:quotient-poset-descent} uses the one-color four-vertex path \(F\) as an example. Merging its endpoints gives a triangle \(Q_1\). Merging one of the two pairs of vertices at distance two gives \(Q_2\), a three-vertex path with one double and one single edge; two label-coincidence patterns produce this isomorphism class. Merging the two nonadjacent vertices of \(Q_2\) then gives the triple-edge graph \(R\). Alternatively, both pairs can be merged simultaneously in \(F\) to produce \(R\) directly.

The forward relations in this example are

\[
H_R=J_R,\qquad
H_{Q_1}=J_{Q_1},\qquad
H_{Q_2}=J_{Q_2}+J_R,
\]

and

\[
H_F=J_F+J_{Q_1}+2J_{Q_2}+J_R.
\]

The coefficient 2 arises from the two label-coincidence patterns that produce isomorphic copies of \(Q_2\). Following the order at the right of Figure~\ref{fig:quotient-poset-descent} gives

\[
J_R=H_R,\qquad
J_{Q_1}=H_{Q_1},\qquad
J_{Q_2}=H_{Q_2}-J_R,
\]

and finally

\[
J_F=H_F-J_{Q_1}-2J_{Q_2}-J_R.
\]

In the directed acyclic graph, the expansion of \(H_F\) contains only \(F\) and graph patterns reachable downstream from \(F\). Ordering all patterns by increasing vertex count turns

\[
H_F=J_F+\sum_{Q\prec F}c(F,Q)J_Q
\]

into a triangular transformation with unit diagonal. For a graph that admits no further vertex merge, \(J_F=H_F\). Every other graph is evaluated in this order as

\[
J_F=H_F-\sum_{Q\prec F}c(F,Q)J_Q,
\]

where all \(J_Q\) on the right have already been obtained. Once the dual tensor networks supply all \(H_F\), an upward recursion from graphs with fewer vertices recovers all \(J_F\). Division by the order of the color-preserving automorphism group then gives

\[
P_F=\frac{J_F}{|\operatorname{Aut}_{\chi}(F)|}.
\]

The entailment relations and integer coefficients \(c(F,Q)\) depend only on graph patterns and can be generated offline before evaluation, independently of molecular size and geometry. Combinatorially, this descent expresses M\"obius inversion on the lattice of set partitions within each vertex color class in terms of graph patterns. Combining partitions that produce isomorphic quotient graphs yields the triangular relation above with integer multiplicities. Here this inversion is specialized to graph homomorphisms with vertex colors, edge multiplicities, and numerical edge weights \cite{Lovasz1967Structures,Lovasz2012GraphLimits,Curticapean2017Homomorphisms}.

This descent does not require an entire degree layer to be evaluated at once. If a selected pattern set \(\mathcal S\) is downward closed under quotient operations, so that \(F\in\mathcal S\) and \(Q\preceq F\) imply \(Q\in\mathcal S\), then evaluating only \(\{H_F:F\in\mathcal S\}\) is sufficient to recover every \(\{P_F:F\in\mathcal S\}\) within the set. Any quotient-closed subset can therefore be evaluated independently.

For a set that is not closed, the required quotient graphs can be added automatically as computational intermediates without exposing them as final descriptors. This allows patterns to be batched according to treewidth, memory, or model selection while preserving the completeness of the algebraic transformation.

\subsubsection{3.5 Variable elimination, treewidth, and computational complexity}\label{sec:treewidth-complexity}

For disconnected \(F\), \(H_F\) is simply the product of the scalar contraction values of its connected components. Once all low-degree connected components have been evaluated, a disconnected pattern therefore costs only a trivial amount. The primary complexity lies in evaluating connected components.

Changing summation order can reduce the cost of a connected \(H_F\) with \(N\) indices well below \(\mathcal O(M^N)\). For example, the homomorphism sum corresponding to Figure~\ref{fig:monomial-to-cmg-ammonia}(c) satisfies

\[
\sum_{i=1}^{M_N}\sum_{j=1}^{M_H}\sum_{k=1}^{M_H}
r_{ij}^{2}r_{jk}
=
\sum_{j=1}^{M_H}
\left(\sum_{i=1}^{M_N}r_{ij}^{2}\right)
\left(\sum_{k=1}^{M_H}r_{jk}\right).
\]

The triple loop on the left requires approximately \(\mathcal O(M_NM_H^2)\) operations. On the right, computing the two row sums separately, multiplying them elementwise, and summing requires only \(\mathcal O(M_HM_N+M_H^2)\) operations. When the element counts have the same scale \(M\), the complexity falls from \(\mathcal O(M^3)\) to \(\mathcal O(M^2)\).

Computing \(H_F\) is equivalent to variable elimination in a graphical model, or contraction of a pairwise factor network \cite{Dechter1999BucketElimination,MarkovShi2008TensorNetworks}. Choose a vertex-elimination order. When eliminating a vertex \(v\), multiply all current factors containing its index, sum over that index, and make the uneliminated neighbors of \(v\) into a clique. The largest number of neighbors encountered during elimination is the induced width of that order. The minimum induced width over all orders equals the treewidth of the support graph,

\[
w(F)=\operatorname{tw}(\operatorname{supp}F).
\]

If every element count is bounded by \(M=\max_a M_a\), the optimal order has the time-complexity bound

\[
\mathcal O\!\left(M^{w(F)+1}\right),
\]

up to graph-pattern-dependent constants. The maximum number of indices in an intermediate tensor is at most \(w(F)\) or \(w(F)+1\), depending on whether the currently eliminated index is included in the reported tensor rank.

This complexity depends on the treewidth of the support graph, not on its number of vertices. Trees, paths, and arbitrary stars all have treewidth 1, so their evaluation is at most quadratic in system size; only dense clique-like structures increase the exponent.

Let \(d\) be the total edge multiplicity of the multigraph. Its support graph has at most \(d\) edges. 
Our exhaustive computational verification shows that\cite{Arnborg1990Partial3Trees,McKayPiperno2014Nauty,McKayGraphData}, 
for \(d\le14\), the maximum treewidth attainable by a simple support graph with at most \(d\) edges is

\[
\boxed{
w_{\max}(d)
=
\left\lfloor
\frac{\sqrt{1+8d}-1}{2}
\right\rfloor ,
\qquad 1\le d\le14 .
}
\]

The corresponding worst-case evaluation exponent is \(M^{w_{\max}(d)+1}\). For example,

\begin{longtable}[]{@{}lrrrr@{}}
\toprule\noalign{}
Algebraic degree \(d\) & \(1\!-\!2\) & \(3\!-\!5\) & \(6\!-\!9\) & \(10\!-\!14\) \\
\midrule\noalign{}
\endhead
\bottomrule\noalign{}
\endlastfoot
Maximum treewidth \(w_{\max}\) & 1 & 2 & 3 & 4 \\
Maximum system-size exponent & \(M^2\) & \(M^3\) & \(M^4\) & \(M^5\) \\
\end{longtable}

Within this degree range, the maximum treewidth is attained by the largest complete graph permitted by the edge budget. 
However, the treewidth and maximum-clique formulas do not coincide in general; this correspondence holds only at low degree.

\subsubsection{3.6 Prefactors for complete degree layers and reuse across patterns}\label{sec:prefactors}

Treewidth controls only the highest power of system size \(M\) for a single graph pattern. For \(M\) identical atoms, the complexity of evaluating the complete PIP set through degree \(d\) can be written as

\[
T_{\le d}(M)=\mathcal O\!\left(f(d)M^{p(d)}\right),
\qquad
p(d)=w_{\max}(d)+1,
\]

where \(p(d)\) is set by the maximum treewidth in this degree range and \(f(d)\) is the effective prefactor of the highest-width contractions. An optimal shared contraction plan must jointly choose contraction orders for individual graphs, intermediate tensors worth retaining, and common subexpressions across graphs; no general closed form for \(f(d)\) is currently known.

Within the low-order regime considered here, however, \(f(d)\) is generally much smaller than the total number of PIP patterns. First, only a few graph patterns attain the maximum treewidth, while most require lower-exponent computations. For example, there are five connected one-color patterns of exact degree three, but only the triangle \(K_3\) has treewidth 2 and requires an \(M^3\)-scale contraction; every other pattern costs at most \(M^2\). Second, graph patterns of the same or different degrees often share partial contractions, so intermediate tensors can be reused. The highest-order cost of a complete basis is not the sum of the highest-order cost of each individual PIP.

The complete basis through degree five for a one-element system illustrates this clearly. Let \(X\in\mathbb R^{M\times M}\) be the first-order pair-variable matrix and first compute

\[
B=X^2.
\]

The triangle homomorphism sum is obtained from \(\langle X,B\rangle\). Once \(B\) is stored, every fourth-order graph pattern can be evaluated through inner products, elementwise products, row reductions, and products of lower-order results involving \(B\), \(X\), and elementwise powers of \(X\). No additional \(M^3\)-scale matrix contraction is required. Thus,

\[
T_{\le4}(M)=M^3+h(4)M^2+O(M).
\]

In terms of leading-order cost, extending the complete third-order basis to the complete fourth-order basis is therefore ``free'': every new operation is at most \(O(M^2)\).

At fifth order, only two additional \(M^3\)-scale shared intermediates are required beyond \(B\). Define \(Y=X^{\circ2}\) and compute

\[
C=XB=X^3,
\qquad
Z=XY=X(X^{\circ2}).
\]

The matrix \(C\) is used for the five-cycle, since \(H_{C_5}=\langle B,C\rangle\), and \(Z\) is used for patterns containing a weighted triangular core of weights \((2,2,1)\). Every remaining fifth-order graph pattern can be evaluated from \(B\), \(C\), and \(Z\) using operations of at most \(O(M^2)\). The complete basis through degree five therefore requires only three \(M^3\)-scale shared contraction nodes, rather than one maximum-width contraction for every PIP:

\[
T_{\le5}(M)=3M^3+h(5)M^2+O(M).
\]

The coefficient 3 denotes the number of leading-order nodes in this shared contraction DAG. Because matrix multiplication requires multiplications and additions and can sometimes exploit matrix symmetry, it is not a literal floating-point operation count. This example shows that \(f(d)\) is only weakly related to the number of PIPs and much more strongly related to the irreducible leading-order computations in the selected shared contraction plan.

\subsubsection{3.7 Black-and-white coloring and canonical deduplication of open intermediate tensors}\label{sec:open-intermediate-tensors}

The quantities \(B=X^2\), \(C=XB\), and \(Z=XY\) in Section 3.6 are intermediate tensors produced during contraction, rather than final scalar outputs. To represent both closed homomorphism sums and open intermediates in the same graph language, we assign each vertex a black or white identity in addition to its element color. Let the vertices of an element-colored multigraph \(F\) be partitioned into black vertices \(\mathcal B\) and white vertices \(\mathcal W\). The corresponding open tensor is defined as

\[
T_F(\boldsymbol i_{\mathcal B})
=
\sum_{\substack{i_w\in\mathcal A_{\chi(w)}\\w\in\mathcal W}}
\prod_{\substack{\{u,v\}\subseteq V(F)\\m_{uv}>0}}
\left(X^{(\chi(u),\chi(v))}_{i_ui_v}\right)^{m_{uv}}.
\]

\begin{figure}[htbp]
\centering
\begin{minipage}[t]{0.48\linewidth}
  \centering
  \includegraphics[width=0.62\linewidth]{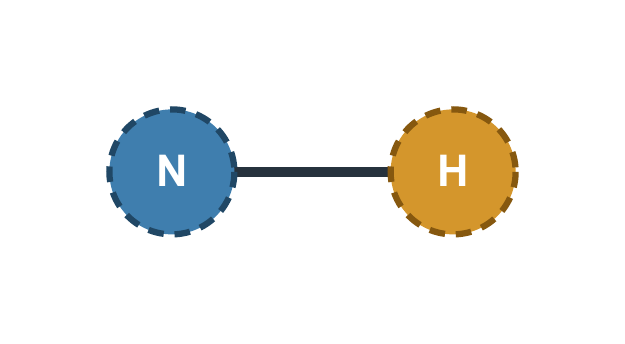}\\[-0.2em]
  \textbf{(a)}
\end{minipage}\hfill
\begin{minipage}[t]{0.48\linewidth}
  \centering
  \includegraphics[width=0.62\linewidth]{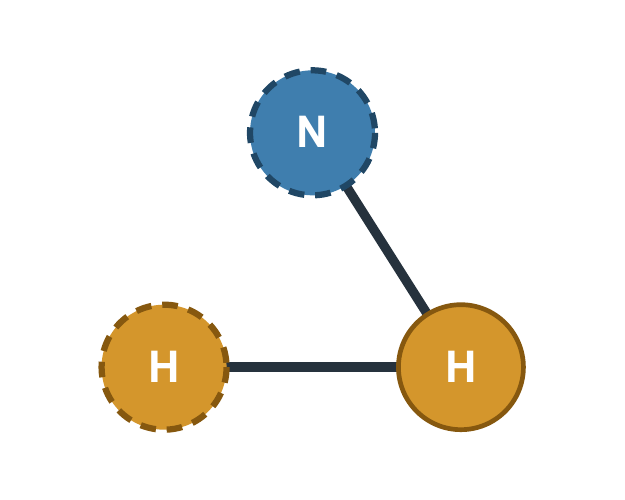}\\[-0.2em]
  \textbf{(b)}
\end{minipage}

\vspace{0.8em}

\begin{minipage}[t]{0.48\linewidth}
  \centering
  \includegraphics[width=0.62\linewidth]{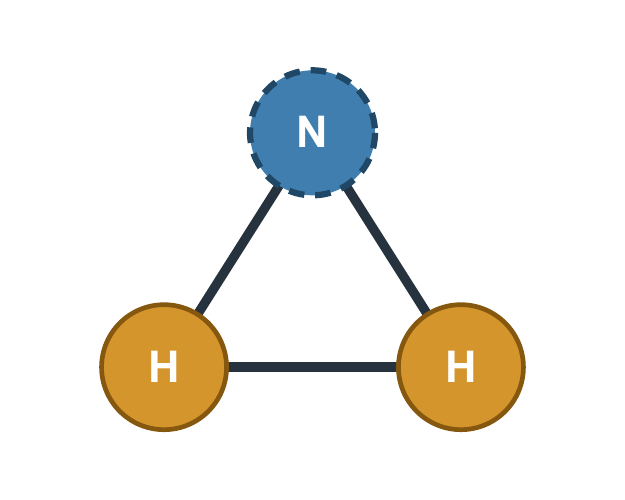}\\[-0.2em]
  \textbf{(c)}
\end{minipage}\hfill
\begin{minipage}[t]{0.48\linewidth}
  \centering
  \includegraphics[width=0.62\linewidth]{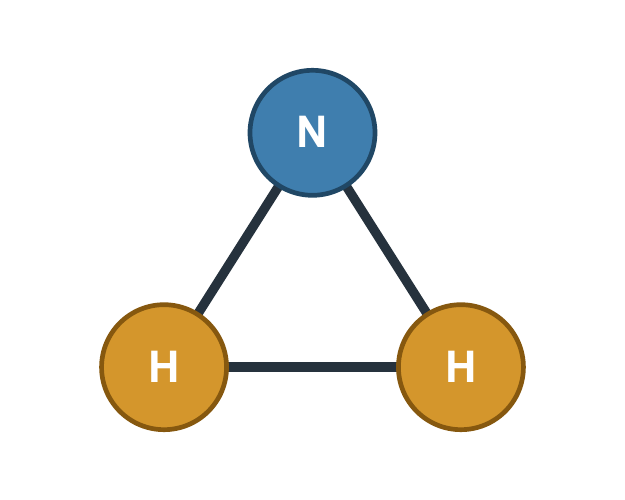}\\[-0.2em]
  \textbf{(d)}
\end{minipage}
\caption{Examples of black-and-white coloring. Dashed outlines denote black vertices and solid outlines denote white vertices; these identities do not denote chemical elements. \textbf{(a)} Two black vertices represent the two free indices of a pair variable. \textbf{(b)} Matrix multiplication sums over the intermediate H index. \textbf{(c)} Adding the third edge to (b) and eliminating another H index. \textbf{(d)} Eliminating the N index from (c) gives a closed scalar with no free indices.}
\label{fig:open-tensor-black-white-contraction}
\end{figure}

The black-vertex indices \(\boldsymbol i_{\mathcal B}\) are retained and are not summed in this expression; they are the free indices of the output tensor. White-vertex indices are summed over their respective element domains and are dummy indices internal to the contraction. The number of black vertices \(|\mathcal B|\) is the rank of \(T_F\). When \(\mathcal B=\varnothing\), the open tensor reduces to the closed homomorphism sum \(H_F\) defined in Section 3.2, as in Figure~\ref{fig:open-tensor-black-white-contraction}(d). Conversely, a pair-variable matrix is an open graph with two black vertices and no white vertices, as in Figure~\ref{fig:open-tensor-black-white-contraction}(a).

Figure~\ref{fig:open-tensor-black-white-contraction}(b) shows the matrix multiplication associated with a typical \(M^3\)-scale pattern. The endpoints \(i,j\) remain black, while the intermediate index \(k\) has been summed and is therefore white.

\[
B_{ij}
=
\sum_{k\in\mathcal A_{\mathrm H}}
X^{(\mathrm{H,H})}_{ik}X^{(\mathrm{H,N})}_{kj}.
\]

Multiplying by the missing edge of the triangle and summing over \(i\) gives the pattern in Figure~\ref{fig:open-tensor-black-white-contraction}(c). Only \(j\) remains free: the N vertex is black and the two H vertices are white.

\[
C_j
=
\sum_{i\in\mathcal A_{\mathrm H}}
X^{(\mathrm{H,N})}_{ij}B_{ij}
=
\sum_{i,k\in\mathcal A_{\mathrm H}}
X^{(\mathrm{H,H})}_{ik}
X^{(\mathrm{H,N})}_{kj}
X^{(\mathrm{H,N})}_{ij}.
\]

Finally, summing over \(j\) gives Figure~\ref{fig:open-tensor-black-white-contraction}(d).

\[
H_F
=
\sum_{j\in\mathcal A_{\mathrm N}}C_j.
\]

This representation also provides a direct way to reuse intermediate tensors across graph patterns. Open graphs are canonicalized while preserving element colors, black-and-white identities, and edge multiplicities. Two open graphs with the same canonical form represent the same intermediate tensor, up to a permutation of its free-index axes. Tensor-axis order must be recorded separately: when the same canonical open graph occurs with a different ordering of its black vertices, only the corresponding axis permutation needs to be aligned; no new tensor value needs to be constructed or evaluated. A canonical open graph therefore needs to be stored only once in the shared contraction DAG and can be reused across contractions of different patterns and degrees.

Given a set of graph patterns and their elimination orders, a shared contraction DAG can be constructed incrementally by checking whether the canonical open graph of each new intermediate already exists. This removes computations that have already appeared. The present heuristic does not, however, guarantee a globally optimal shared contraction plan. Different elimination-order algorithms may produce different sets of intermediates and thereby affect \(f(d)\) in Section 3.6.

\subsubsection{3.8 Contraction-order heuristics and sloped truncation}\label{sec:sloped-truncation}

Finding an exact optimal tree decomposition is itself computationally difficult. We use a min-fill/min-degree heuristic to generate elimination orders \cite{Bodlaender2010Treewidth}. At each step, for every candidate vertex we count the edges that must be added to make its current neighbors a clique after elimination. We select the vertex requiring the fewest added edges; ties are broken by the smallest current degree and then by canonical vertex order. Both degree and fill count are measured on the current graph after all previous fill edges, not on the original support graph.

Complete enumeration shows that, for every nonisomorphic support graph with \(d\le12\), every min-fill/min-degree tie path reaches the exact treewidth. A nonoptimal order does not change the contraction value, but it increases intermediate-tensor sizes and computational cost.

In addition to the elimination order, we jointly control a complete PIP basis by algebraic degree, number of connected components, and maximum treewidth along the descent path. These three quantities form a discrete three-dimensional truncation surface, which we call sloped truncation.

For a graph pattern \(F\), define its maximum descent-path treewidth as

\[
w_{\mathrm{path}}(F)
=
\max_{Q\preceq F}
\operatorname{tw}\!\left(\operatorname{supp}Q\right),
\]

where \(Q\preceq F\) ranges over all quotient graphs obtainable by merging same-color nonadjacent vertices, including \(F\) itself. Given a maximum degree \(d_{\max}\), a degree-dependent component cutoff \(c_{\mathrm{cut}}(d)\), and a degree-dependent path-treewidth cutoff \(w_{\mathrm{cut}}(d)\), the retained pattern set is

\[
\mathcal B(d_{\max},c_{\mathrm{cut}},w_{\mathrm{cut}})
=
\left\{
F:
\deg(F)\le d_{\max},\
\kappa(F)\le c_{\mathrm{cut}}(\deg F),\
w_{\mathrm{path}}(F)\le w_{\mathrm{cut}}(\deg F)
\right\}.
\]

The three cutoffs play distinct roles. Algebraic degree controls the maximum polynomial order. The number of connected components limits the number of disconnected patterns produced by products of lower-order connected patterns. Maximum path treewidth limits the greatest contraction width needed both for the homomorphism sums and for the complete descent. The maximum must be taken over the entire descent closure, rather than from \(F\) alone, because recovering \(P_F\) requires all quotient graphs \(Q\preceq F\). Quotient operations preserve algebraic degree and cannot increase the number of connected components, while every \(Q\preceq F\) satisfies \(w_{\mathrm{path}}(Q)\le w_{\mathrm{path}}(F)\). Hence, whenever \(F\) passes the sloped truncation, every quotient required by its descent passes the same degree-layer cutoff, and the retained set is downward closed under quotient operations.

Compared with truncation by degree alone, sloped truncation can retain higher-degree star-, chain-, or tree-like patterns with few connected components and low contraction width, while removing patterns with too many disconnected combinations or expensive descent contractions. It therefore directly controls the tradeoff among polynomial degree, basis size, and numerical evaluation cost.

\subsubsection{3.9 Automatic differentiation and force-label training}\label{sec:automatic-differentiation}

The entire evaluation pipeline consists only of elementwise matrix functions, multiplication, summation, tensor contraction, and linear combinations with precomputed coefficients. It can therefore be implemented directly in automatic-differentiation frameworks such as PyTorch and JAX. If a model predicts an energy \(E_\theta(\mathbf R)\), the atomic forces are

\[
\mathbf F_i=-\frac{\partial E_\theta}{\partial\mathbf R_i},
\]

which guarantees energy--force consistency. Training against force labels requires mixed second derivatives of the energy with respect to coordinates and parameters; modern reverse-mode automatic differentiation can evaluate them directly \cite{Baydin2018AD}. Previous PIP work likewise shows that analytic gradients or reverse differentiation are important for efficient force training \cite{Nandi2019PIPGradients,Houston2022PIPGradients}. The effects of force labels on model accuracy have also been studied systematically in the learning of molecular energies and forces \cite{Christensen2020Gradients}.

The key advantage of the present method is that its computational graph remains compact rather than expanding into a potentially enormous monomial list. M\"obius-transform coefficients and graph canonicalization are coordinate independent and do not participate in differentiation; reverse propagation passes only through numerical tensor contractions. Energies and forces can therefore share the same PIP definitions without separately maintained symbolic-derivative code.

\subsection{4 Transferable CMG-PIP descriptors across systems}\label{sec:transferable-descriptors}

\begin{figure}[htbp]
\centering
\begin{tikzpicture}[
  panel/.style={inner sep=0pt,outer sep=0pt},
  flow/.style={-{Latex[length=2.4mm,width=1.9mm]},draw=black!65,line width=0.75pt},
  flowlabel/.style={align=center,text=black!65,inner sep=1pt}
]
\node[panel] (figfoura) {\includegraphics[width=0.23\linewidth]{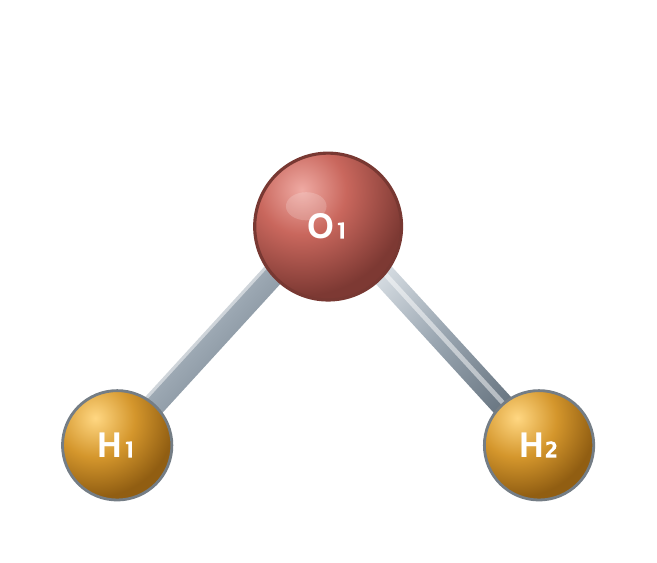}};
\node[panel,right=0.155\linewidth of figfoura] (figfourb) {\includegraphics[width=0.23\linewidth]{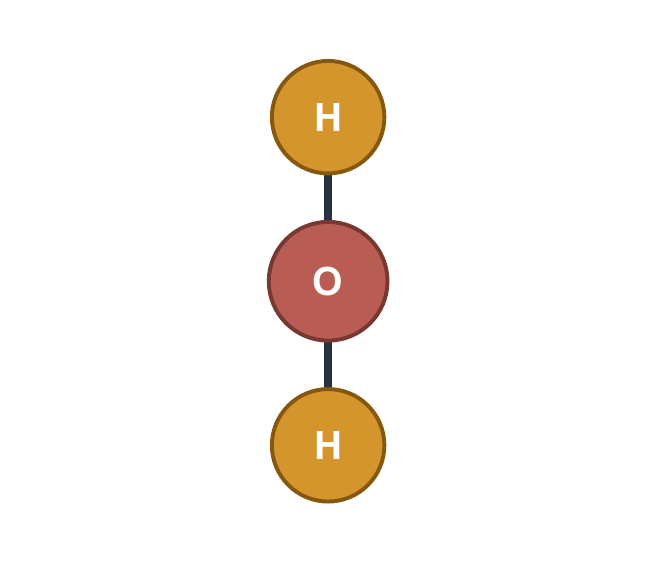}};
\node[panel,right=0.155\linewidth of figfourb] (figfourc) {\includegraphics[width=0.23\linewidth]{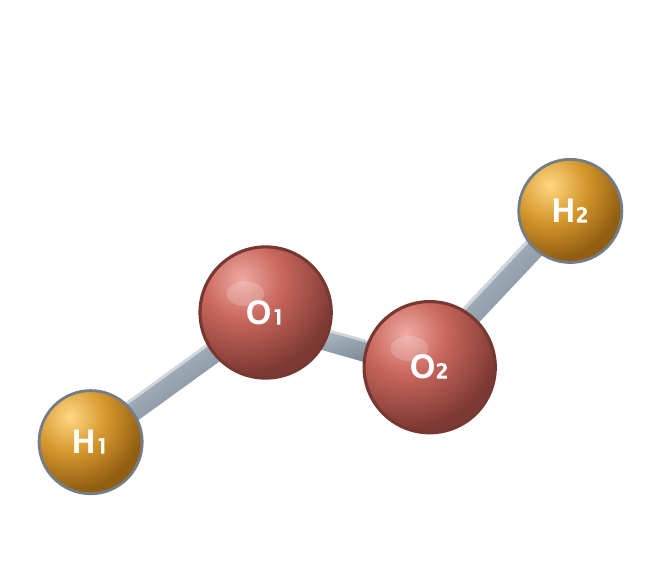}};

\node[anchor=south west,font=\bfseries\small] at (figfoura.north west) {(a)};
\node[anchor=south west,font=\bfseries\small] at (figfourb.north west) {(b)};
\node[anchor=south west,font=\bfseries\small] at (figfourc.north west) {(c)};

\draw[flow] (figfourb.west) -- (figfoura.east);
\node[flowlabel] at ($(figfourb.west)!0.5!(figfoura.east)+(0,0.43cm)$)
  {\footnotesize instantiate on \(\mathrm{H_2O}\)};

\draw[flow] (figfourb.east) -- (figfourc.west);
\node[flowlabel] at ($(figfourb.east)!0.5!(figfourc.west)+(0,0.43cm)$)
  {\footnotesize instantiate on \(\mathrm{H_2O_2}\)};
\end{tikzpicture}
\caption{Instantiation of the same colored multigraph in different molecular systems. \textbf{(a)} A water molecule with atom labels. \textbf{(b)} A general graph pattern \(F\) without concrete atom labels; it is labeled and summed separately over the atom sets of water and hydrogen peroxide. \textbf{(c)} A hydrogen-peroxide molecule with atom labels. The two instantiations produce \(P_F^{(\mathrm{H_2O})}\) and \(P_F^{(\mathrm{H_2O_2})}\), which belong to different variable spaces. Light-gray lines in the molecular sketches denote chemical bonds only for molecular identification and are not edges of the graph pattern \(F\).}
\label{fig:shared-cmg-water-peroxide}
\end{figure}

Constructing a transferable potential from PIP descriptors encounters a theoretical difficulty: different molecular systems have entirely different explicit PIP expressions. The graph representation reveals that the graph pattern itself is invariant across systems. The same colored multigraph can be labeled on molecular systems with different atom counts, producing PIPs in their respective variable spaces. Figure~\ref{fig:shared-cmg-water-peroxide} uses an H---O---H path to show how one graph pattern is instantiated in water and hydrogen peroxide.

In water, after removing the repetition caused by exchanging the two H endpoints, \(F\) gives only one distinct monomial:

\[
P_F^{(\mathrm{H_2O})}
=x_{O_1H_1}x_{O_1H_2}.
\]

In hydrogen peroxide, the central O vertex may be labeled \(O_1\) or \(O_2\), so the corresponding orbit sum is

\[
P_F^{(\mathrm{H_2O_2})}
=x_{O_1H_1}x_{O_1H_2}
+x_{O_2H_1}x_{O_2H_2}.
\]

The two PIPs have different analytic expressions but describe the same graph pattern. This is the necessary basis for a descriptor that is invariant across systems.

This section defines only the descriptors and their symmetries, without restricting the downstream regression model. They can be supplied to linear models, neural networks, or other differentiable functions. Specific network architectures, training data, and loss designs are deferred to application studies.

\subsubsection{4.1 System-independent global graph patterns}\label{sec:global-patterns}

Given an element alphabet \(\mathcal C\), a maximum degree \(d_{\max}\), and an optional contraction-width bound \(p_{\max}\), all \(|\mathcal C|\)-colored graph patterns satisfying the truncation can be enumerated once. For any particular molecule, color-exhausted patterns are set to zero and \(P_F\) is evaluated for the remaining patterns. All molecules using the same element alphabet and truncation rules therefore share a descriptor vector of fixed length and fixed semantics.

To reduce the purely combinatorial growth of a pattern with system size, we can normalize by the number of atom subsets selected in the first stage:

\[
\widetilde P_F
=
\frac{P_F}
{\displaystyle\prod_a\binom{M_a}{n_a(F)}} ,
\]

When the denominator is nonzero, this is the average of the local orbit sums over all atom subsets satisfying the color counts. For a color-exhausted pattern, we define \(\widetilde P_F=0\). This normalization preserves permutation invariance and the M\"obius coefficients between patterns; it separates the system-size effect of ``how many atom subsets can be selected'' from the descriptor value.

If pairwise radial functions tend to zero as \(r\rightarrow\infty\), adding an atom infinitely separated from every other atom leaves graph terms that do not involve it unchanged, but changes the combinatorial normalization denominator above. Size transferability of global descriptors therefore does not automatically imply strict size extensivity. For potentials of large systems, a more natural approach is usually to use the local centered descriptors below and sum local energies. Global descriptors are suited to total molecular properties or datasets spanning a limited range of system sizes.

\subsubsection{4.2 The zeroth element and atom-centered descriptors}\label{sec:zero-element-atom-center}

Consider a specified atom \(i\) and the sum of pair variables involving the surrounding hydrogen atoms:

\[
d_i^{(\mathrm H)}
=\sum_{j\in\mathcal A_{\mathrm H}\setminus\{i\}}x_{ij}.
\]

A corresponding two-leaf polynomial can also be constructed:

\[
\sum_{\substack{j<k\\j,k\in\mathcal A_{\mathrm H}\setminus\{i\}}}x_{ij}x_{ik}
\]

\begin{figure}[htbp]
\centering
\begin{minipage}[c]{0.46\linewidth}
  \centering
  \includegraphics[width=0.9\linewidth]{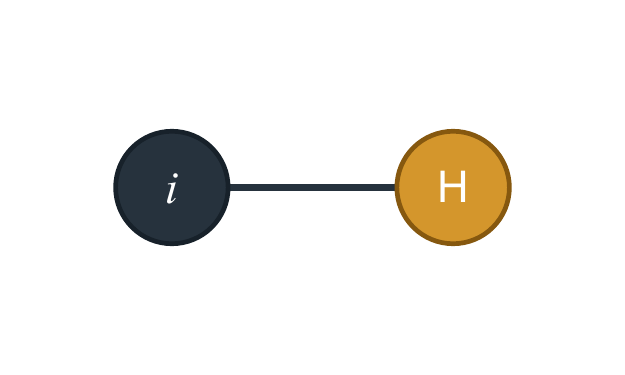}
  \par\smallskip\textbf{(a)}
\end{minipage}\hfill
\begin{minipage}[c]{0.46\linewidth}
  \centering
  \includegraphics[width=0.9\linewidth]{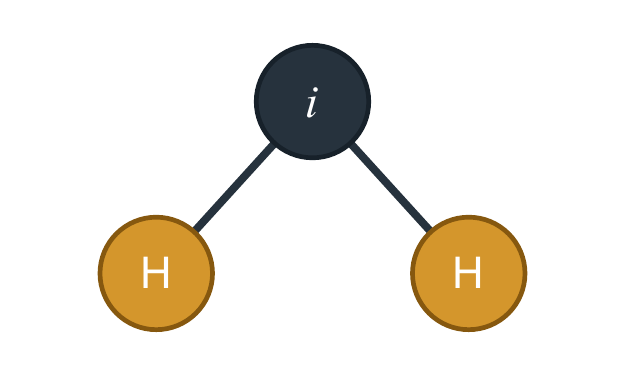}
  \par\smallskip\textbf{(b)}
\end{minipage}
\caption{Graph representations of atom-centered polynomials. The black vertex is the fixed center \(i\), and yellow vertices are environmental H atoms. \textbf{(a)} A single-edge pattern. \textbf{(b)} A two-leaf pattern. The two graphs correspond to the two polynomials above.}
\label{fig:atom-center-patterns}
\end{figure}

Both expressions are invariant under permutations of like atoms outside the center. They can therefore be regarded as permutationally invariant polynomials centered on atom \(i\), and represented by colored multigraphs as in Figure~\ref{fig:atom-center-patterns}.

We introduce a special color for the center, which may be called the center color, the zeroth-element color, or an additional selected color. For convenience, we use black. A colored multigraph with exactly one vertex of this center color is called atom-centered.

\begin{figure}[htbp]
\centering
\includegraphics[width=0.86\linewidth]{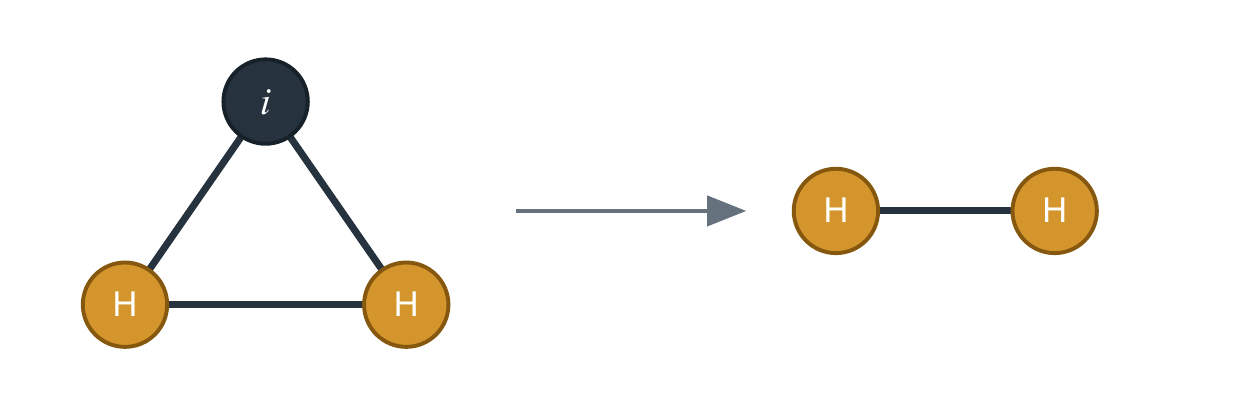}
\par\smallskip\textbf{(a)}
\par\medskip
\includegraphics[width=0.86\linewidth]{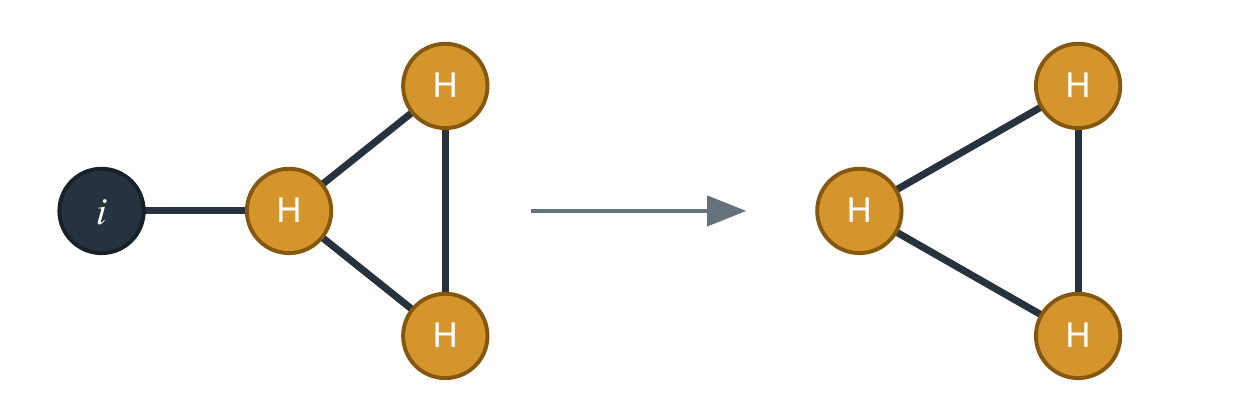}
\par\smallskip\textbf{(b)}
\caption{Contraction graphs with a fixed center. The black vertex is the fixed center \(i\), yellow vertices are environmental H atoms, and the right-hand graphs are the environmental induced graphs after the center is removed. \textbf{(a)} The center lies on the triangle, leaving a graph of treewidth 1. \textbf{(b)} The center is an attached leaf, leaving treewidth 2.}
\label{fig:atom-center-treewidth}
\end{figure}

The actual element of the center does not change the form of the PIP or its graph pattern. Different central elements can instead be distinguished through the weights of a subsequent neural network or an additional element label, rather than through different center colors. This additional center coloring leaves the procedure for computing graph-isomorphism coefficients in the descent unchanged, but affects contraction complexity. The complexity is controlled by the maximum treewidth of the connected components remaining after the center vertex is removed. For example,

Figure~\ref{fig:atom-center-treewidth}(a) fixes a vertex of a triangle as the center, leaving two summation indices; the homomorphism sum for one center can be evaluated in \(O(M^2)\). In Figure~\ref{fig:atom-center-treewidth}(b), the center is a leaf attached to the triangle. The environmental triangle still requires contraction, so dense evaluation for one center remains \(O(M^3)\). Here \(M\) is the number of atoms in the system.

Evaluating these descriptors separately for each of the \(M\) atoms increases the total online complexity. For example, evaluation through degree five would rise from \(O(M^3)\) to \(O(M^4)\). Batched evaluation addresses this problem.

\subsubsection{4.3 Black-and-white coloring and batched atom-centered evaluation}\label{sec:atom-center-batch}

We have used the same colored multigraph \(F\) to represent its associated \(H_F\), \(J_F\), and \(P_F\). In \hyperref[sec:open-intermediate-tensors]{Section 3.7}, we introduced a parallel black-and-white coloring while retaining the element colors of both black and white vertices.
There, black and white vertices represent free and dummy indices. A central atom likewise plays the role of a free index, especially when polynomials such as that in Figure~\ref{fig:atom-center-treewidth}(a) are assembled into a one-dimensional tensor over all centers. The same black-and-white coloring can therefore be used for atom-centered representations. Intermediate tensors with exactly one black vertex in \hyperref[sec:open-intermediate-tensors]{Section 3.7} can be interpreted as \(\vec H_F\) for a set of atom-centered PIPs. The arrow denotes a vector containing \(H_F\) for all central atoms, with \(H_F(i)\) the component for center \(i\). We use the same notation for \(\vec J_F\) and \(\vec P_F\).

Descent from these \(\vec H_F\) evaluates all corresponding \(\vec J_F\) in a batch. The rules differ slightly from both the all-white global representation and the fixed-center representation in \hyperref[sec:zero-element-atom-center]{Section 4.2}:

Black-and-white identity does not enter the test for equality of element colors. Only nonadjacent vertices with the same element color are merged. If either merged vertex is black, the resulting vertex is black. Label partitions are still generated over all vertices of each element, and descent becomes a linear operation on vectors rather than scalars.

\[
\vec J_F
=\sum_{\pi\in\prod_a\Pi(V_a(F))}
\mu(\hat 0,\pi)\,\vec H_{F/\pi}.
\]
Here \(V_a(F)\) includes both black and white vertices of element \(a\), and \(\mu\) is the partition-lattice M\"obius coefficient defined above. Terms that produce loops vanish because \(x_{ii}=0\).

\begin{figure}[htbp]
\centering
\includegraphics[width=\linewidth]{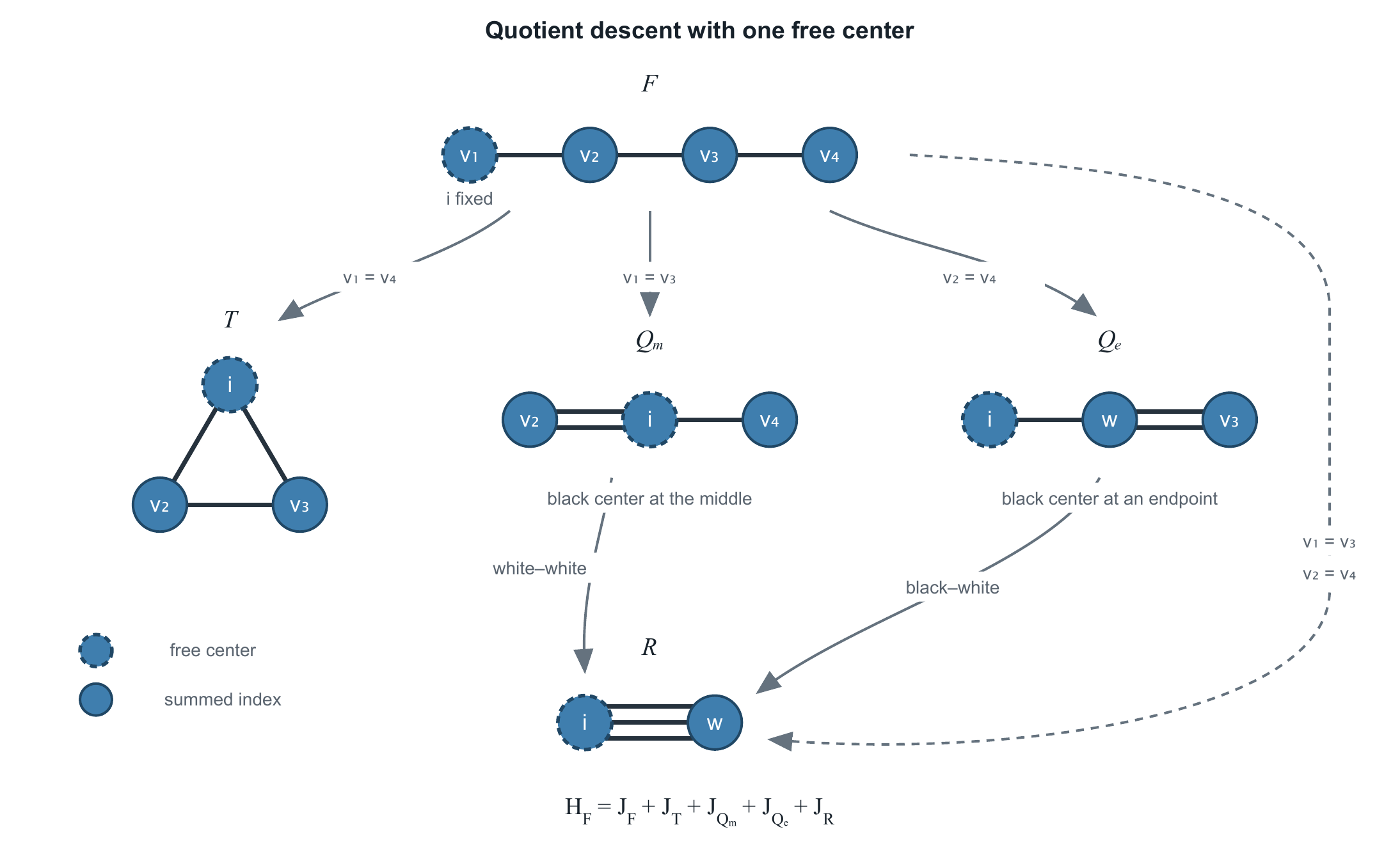}
\caption{Atom-centered descent for a four-vertex path. Dashed outlines mark the free center index. Solid arrows denote one vertex coincidence, and dashed arrows denote simultaneous coincidences of two vertex pairs. \(Q_{\mathrm m}\) and \(Q_{\mathrm e}\) are distinguished by the black-vertex position; \(R\) is the triple-edge graph.}
\label{fig:atom-center-rooted-descent}
\end{figure}

Figure~\ref{fig:atom-center-rooted-descent} gives an example analogous to Figure~\ref{fig:quotient-poset-descent}, with \(v_1\) black and the other vertices white, all belonging to the same element. In each component of \(\vec H_F\), the white vertex \(v_4\) may take the center label; merging \(v_1=v_4\) produces the centered triangle \(T\). Merging \(v_1=v_3\) or \(v_2=v_4\) produces \(Q_{\mathrm m}\) or \(Q_{\mathrm e}\), respectively: the black vertex lies in the middle in the former and at an endpoint in the latter. Their unrooted topologies coincide, but their center positions differ, so they must be evaluated as two distinct graph patterns.

The corresponding vector relations are

\[
\vec H_F=\vec J_F+\vec J_T+\vec J_{Q_{\mathrm m}}+\vec J_{Q_{\mathrm e}}+\vec J_R,
\]

\[
\begin{aligned}
\vec H_T&=\vec J_T,\qquad \vec H_R=\vec J_R,\\
\vec H_{Q_{\mathrm m}}&=\vec J_{Q_{\mathrm m}}+\vec J_R,\\
\vec H_{Q_{\mathrm e}}&=\vec J_{Q_{\mathrm e}}+\vec J_R.
\end{aligned}
\]

Descent along the quotient relations gives

\[
\vec J_F=\vec H_F-\vec H_T-\vec H_{Q_{\mathrm m}}-\vec H_{Q_{\mathrm e}}+\vec H_R.
\]

Without black-and-white labels, \(Q_{\mathrm m}\) and \(Q_{\mathrm e}\) combine into a single term with coefficient 2. Retaining the center splits this into two terms with coefficient 1. The black-vertex position therefore enters both quotient-graph deduplication and descent-coefficient computation.

The optimal elimination width when one center index is retained is the same as that of the all-white closed graph. The cost of evaluating the complete \(\vec P_F\) is controlled by the maximum width in its quotient-descent closure. For a fixed number of elements, the complete atom-centered graph basis through degree five can be evaluated in a batch in \(O(f_{\mathrm B}(5,E)M^3)\). Automorphism coefficients must account for the distinction between black and white labels.

\subsubsection{4.4 Atom-pair-centered descriptors}\label{sec:pair-center}

We next consider descriptors centered on two atoms. The atom-pair-centered descriptor set includes both the expected two-black-vertex graphs and two sets of one-black-vertex graphs. The latter can be evaluated using the preceding representation; we discuss only the two-black-vertex part below.

For a specified atom pair \((i,j)\), we can use the center coloring of \hyperref[sec:zero-element-atom-center]{Section 4.2}. Two centers of the same element share a center color with domain \(\{i,j\}\). Centers of different elements use two center colors, each with a singleton domain.

In batched evaluation, the two black vertices retain two free indices, giving a matrix with components \(H_F(i,j)\); descent becomes a linear operation on matrices. Black vertices of different elements cannot take the same atom label, so the rules of \hyperref[sec:atom-center-batch]{Section 4.3} apply directly. For centers of the same element, the two black vertices cannot coincide in off-diagonal components either. Descent therefore retains only partitions in which the two black vertices belong to different blocks. White--white and black--white merges follow the same rules as in \hyperref[sec:atom-center-batch]{Section 4.3}.

Terms in which the two black vertices coincide occur only on the diagonal. Writing these vertices as \(b_1,b_2\), we have

\[
H_F(i,i)=H_{F/(b_1=b_2)}(i).
\]

The diagonal of the two-center matrix is thus the one-center homomorphism sum obtained by merging the two black vertices. Edge multiplicities are retained; if the black vertices are adjacent, the merge produces a loop and the term vanishes. These quotient graphs can be assigned to the one-center part and descended separately, without subtraction from the off-diagonal atom-pair components. The pair descent therefore never needs to consider coincidences of its black vertices.

For an unordered pair of like atoms, the two center labels have two assignments. Corresponding upper- and lower-triangular components should therefore be added, giving \(H_F(i,j)+H_F(j,i)\), and stored for \(i<j\). The descended \(J\) values are summed in the same way and normalized by the corresponding automorphism coefficient to obtain \(P\).

For molecular centers, membership of the central atoms in the same molecule also enters the definition. Consider a system of \(N\) water molecules. Evaluation for one specified water molecule can still use \hyperref[sec:zero-element-atom-center]{Section 4.2}, with center colors marking the one, two, or three vertices involved in the description. In batched evaluation, however, three black vertices retain three independent atom indices, producing a rank-three tensor with \(O(N^3)\) candidate combinations, most drawn from different water molecules. Only \(N\) molecular centers are actually needed. Extension to molecular centers therefore also requires binding these atom indices by molecular membership, so that one molecular index jointly determines all central atoms.

\subsection{5 Edge coloring and clique tensors}\label{sec:edge-colors-and-clique-tensors}

\subsubsection{5.1 Edge coloring and mixed radial bases}\label{sec:edge-colored-radial-basis}

In practice, PIPs are usually polynomials in scalar functions of bond lengths rather than in bond lengths themselves. Common choices include inverse, Morse-type, and exponential functions, often with specified hyperparameters. These hyperparameters can in principle be optimized. This has rarely been done largely because of limitations of the computational tools: conventional explicit polynomials are difficult to integrate with modern machine-learning frameworks, making it difficult to include such hyperparameters in the differentiation graph and optimize them jointly with other parameters. One strict requirement is that atom pairs belonging to the same function channel and having the same endpoint element types must share the function form and hyperparameters.

Nevertheless, multiple scalar functions of the same type of bond length can be used simultaneously without breaking exact permutation invariance. For example, a triangle with three vertices of the same color can represent the following invariant polynomial:

\[
\sum_i\sum_{j\neq i}\sum_{k\neq i,j}
\frac{1}{r_{ij}}e^{-\alpha r_{jk}}e^{-\beta r_{ki}^{2}}.
\]

This expression uses three different radial-function matrices while remaining exactly permutation invariant. Such a construction may help fitting; at least, it is allowed by symmetry\cite{hao2026SRI}.

For these cases, edge colors can be introduced in addition to vertex colors. Let \(\chi(v)\) denote the vertex color and \(\kappa(e)\) the edge color. The function represented by an edge is jointly determined by its unordered endpoint-color pair and its edge color:

\[
\bigl(\{\chi(u),\chi(v)\},\kappa(e)\bigr)
\]

\begin{figure}[htbp]
\centering
\begin{minipage}[c]{0.46\linewidth}
  \centering
  \includegraphics[width=\linewidth]{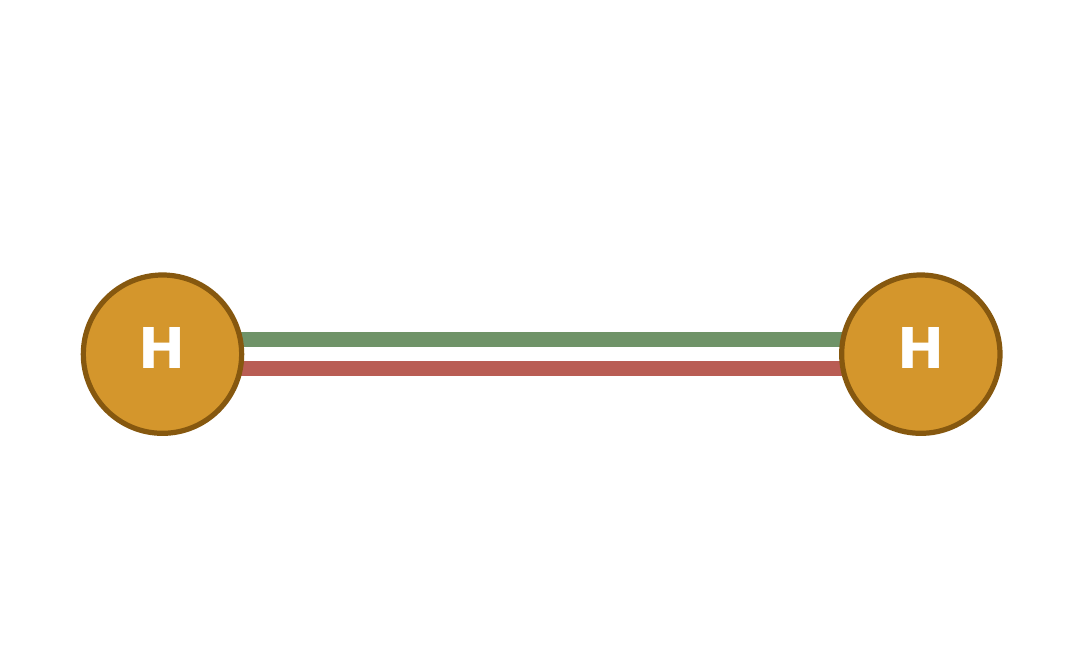}
  \par\smallskip\textbf{(a)}
\end{minipage}
\hfill
\begin{minipage}[c]{0.46\linewidth}
  \centering
  \includegraphics[width=\linewidth]{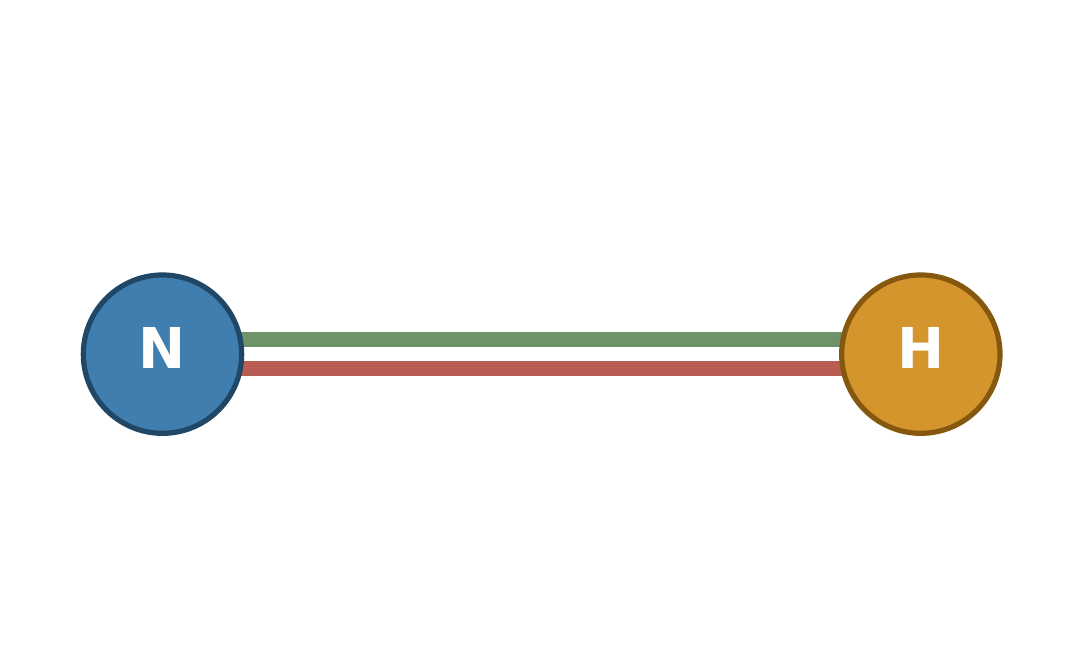}
  \par\smallskip\textbf{(b)}
\end{minipage}
\caption{The same edge colors with different endpoint-color pairs. Two parallel edges denote two radial channels on the same atom pair. \textbf{(a)} An H--H endpoint-color pair. \textbf{(b)} The same two edge colors with an N--H endpoint-color pair. The function of each edge is jointly determined by its unordered endpoint-color pair and its edge color.}
\label{fig:edge-colored-multiradial}
\end{figure}

Parallel edges between the same vertices may carry different colors and hence use different radial functions. Conversely, the same edge color need not represent the same function form when its endpoint elements differ. Figures~\ref{fig:edge-colored-multiradial}(a) and \ref{fig:edge-colored-multiradial}(b) illustrate this case: both panels use exactly the same two edge colors, but their endpoint-color pairs are H--H and N--H, so the functions represented by the two edges may differ.

Figure~\ref{fig:edge-colored-multiradial}(a) can represent the polynomial

\[
\sum^{M_H}_{i}\sum^{M_H}_{j<i}
\frac{1}{r_{ij}}e^{-\alpha r_{ij}}.
\]

The green and coral edges may represent \(r^{-1}\) and \(e^{-\alpha r}\), respectively. Keeping the edge colors unchanged and replacing one endpoint with N gives Figure~\ref{fig:edge-colored-multiradial}(b), which can represent

\[
\sum^{M_N}_{i}\sum^{M_H}_{j}
\frac{1}{r_{ij}}e^{-\beta r_{ij}^{2}}.
\]

Here the green edge may still represent \(r^{-1}\), while the same coral edge on an N--H pair may represent \(e^{-\beta r^2}\). The complete function channel is therefore defined jointly by endpoint colors and edge color.

Edge coloring affects automorphism coefficients and the number of graph patterns, but does not change the underlying uncolored graph. Tensor-contraction complexity can therefore still be estimated from treewidth as before, and descent follows the same principles. Sloped truncation becomes more involved, however, and the number of graph patterns increases.

\subsubsection{5.2 Higher-order internal coordinates and clique tensors}\label{sec:higher-order-internal-coordinates}

PIPs need not be restricted to scalar functions of bond lengths. Other internal coordinates, including bond angles and dihedral angles, can also be incorporated. A permutationally invariant polynomial can contain both bond-length and bond-angle factors, for example:

\[
\sum^{M_O}_{i}\sum^{M_H}_{j}
\sum^{M_H}_{k\neq j}\sum^{M_H}_{l\neq j,k}
\frac{1}{r_{il}}\frac{1}{r_{ij}}\cos\!\left(\theta^i_{jk}\right).
\]

\begin{figure}[htbp]
\centering
\includegraphics[width=0.50\linewidth]{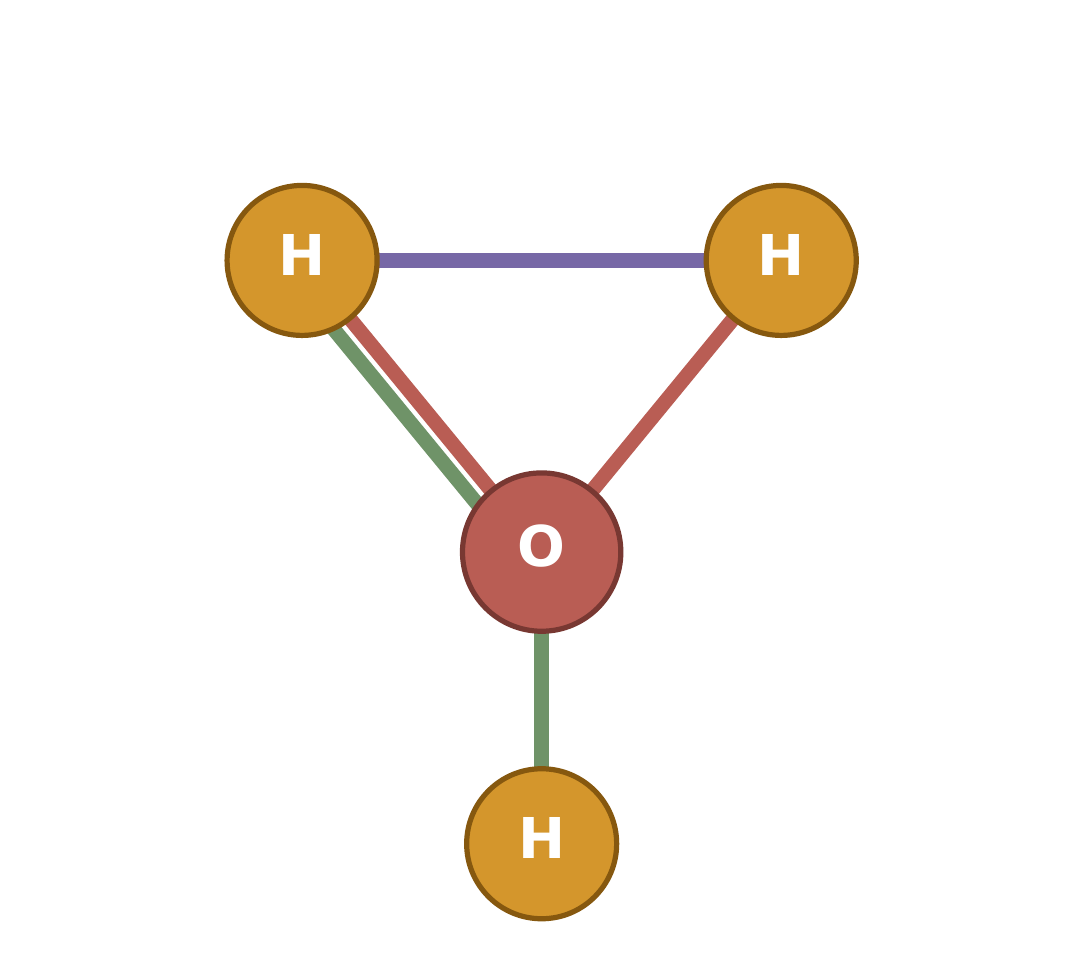}
\caption{A graph pattern containing pairwise radial factors and a rank-three angle tensor. The central vertex is O and the other three vertices are H. Green edges denote the two radial factors. The triangular clique formed by the upper two H vertices and the central O vertex jointly represents the angle tensor, with different edge colors for the two O--H wings and the opposite H--H edge. The left O--H atom pair belongs to both a radial factor and the angle tensor and is therefore drawn with two edges.}
\label{fig:clique-tensor-angle}
\end{figure}

Here \(i\) runs over O atoms and \(j,k,l\) over H atoms; the successive restrictions ensure that the three H labels are distinct. Figure~\ref{fig:clique-tensor-angle} shows the corresponding graph pattern.

A bond angle or a scalar function of it can be regarded as a special rank-three tensor \(A^i_{jk}\), set to zero whenever any two indices coincide and symmetric under exchange of its two wings, \(A^i_{jk}=A^i_{kj}\). Polynomials involving bond angles can still be described in graph language: the factor is represented by a three-vertex clique whose three edges are associated with the same tensor, with different colors for the wings and the opposite edge. These three edges jointly represent one rank-three tensor, rather than three independent pairwise factors. Likewise, a dihedral angle or a scalar function of it can be represented by a four-vertex clique.

Introducing these cliques also affects automorphism coefficients and the number of graph patterns. Higher-order tensors leave the complexity-analysis framework unchanged: after completing the indices of each higher-order factor into a clique, the treewidth of the resulting support graph can still be used to estimate contraction complexity. Descent follows the same principles as before.

\section{Results}\label{sec:results}

\subsection{1 Comparison with previous methods on small systems}\label{sec:small-system-comparison}

We first compare the offline generation time of the CMG graph-pattern method with existing PIP/FI generation programs for several representative permutation groups. All programs are restricted to one thread.
The FI-2025 row reports published results~\cite{Hao2025PIPFIAlgorithms}. FIgenerator is an improved software implementation that retains the overall algorithm of FI-2025, with the main improvement being the replacement of binary search with hash-based lookup. The FIgenerator and CMG data are reruns conducted in this work.
FIgenerator records the CPU time for the target degree alone, whereas CMG records cumulative wall-clock time from process startup through completion of the target degree.

For the \(A_2B_7\) system, the generation times are as follows.

\begingroup
\small
\setlength{\tabcolsep}{4pt}
\begin{longtable}[]{@{}
  >{\raggedright\arraybackslash}p{(\linewidth - 10\tabcolsep) * \real{0.2200}}
  >{\raggedleft\arraybackslash}p{(\linewidth - 10\tabcolsep) * \real{0.1560}}
  >{\raggedleft\arraybackslash}p{(\linewidth - 10\tabcolsep) * \real{0.1560}}
  >{\raggedleft\arraybackslash}p{(\linewidth - 10\tabcolsep) * \real{0.1560}}
  >{\raggedleft\arraybackslash}p{(\linewidth - 10\tabcolsep) * \real{0.1560}}
  >{\raggedleft\arraybackslash}p{(\linewidth - 10\tabcolsep) * \real{0.1560}}@{}}
\toprule\noalign{}
\begin{minipage}[b]{\linewidth}\raggedright
Method
\end{minipage} & \begin{minipage}[b]{\linewidth}\raggedleft
Degrees 1--3
\end{minipage} & \begin{minipage}[b]{\linewidth}\raggedleft
Degree 4
\end{minipage} & \begin{minipage}[b]{\linewidth}\raggedleft
Degree 5
\end{minipage} & \begin{minipage}[b]{\linewidth}\raggedleft
Degree 6
\end{minipage} & \begin{minipage}[b]{\linewidth}\raggedleft
Degree 7
\end{minipage} \\
\midrule\noalign{}
\endhead
\bottomrule\noalign{}
\endlastfoot
FI-2025 & 0.025 s & 0.220 s & 1.500 s & 23.000 s & 344.000 s \\
FIgenerator & 0.000 s & 0.040 s & 0.430 s & 5.190 s & 76.130 s \\
CMG & 0.008 s & 0.054 s & 0.404 s & 3.194 s & 31.887 s \\
\end{longtable}
\endgroup

For the larger permutation groups \(A_3B_9\) and \(A_7B_6C_2\), CMG still generates complete results through degree six in a short time.

\begingroup
\small
\setlength{\tabcolsep}{4pt}
\begin{longtable}[]{@{}
  >{\raggedright\arraybackslash}p{(\linewidth - 10\tabcolsep) * \real{0.1200}}
  >{\raggedright\arraybackslash}p{(\linewidth - 10\tabcolsep) * \real{0.2500}}
  >{\raggedleft\arraybackslash}p{(\linewidth - 10\tabcolsep) * \real{0.1575}}
  >{\raggedleft\arraybackslash}p{(\linewidth - 10\tabcolsep) * \real{0.1575}}
  >{\raggedleft\arraybackslash}p{(\linewidth - 10\tabcolsep) * \real{0.1575}}
  >{\raggedleft\arraybackslash}p{(\linewidth - 10\tabcolsep) * \real{0.1575}}@{}}
\toprule\noalign{}
\begin{minipage}[b]{\linewidth}\raggedright
System
\end{minipage} & \begin{minipage}[b]{\linewidth}\raggedright
Method
\end{minipage} & \begin{minipage}[b]{\linewidth}\raggedleft
Degrees 1--3
\end{minipage} & \begin{minipage}[b]{\linewidth}\raggedleft
Degree 4
\end{minipage} & \begin{minipage}[b]{\linewidth}\raggedleft
Degree 5
\end{minipage} & \begin{minipage}[b]{\linewidth}\raggedleft
Degree 6
\end{minipage} \\
\midrule\noalign{}
\endhead
\bottomrule\noalign{}
\endlastfoot
\(A_3B_9\) & FIgenerator & 0.020 s & 0.500 s & 10.730 s & 213.840 s \\
\(A_3B_9\) & CMG & 0.010 s & 0.096 s & 0.962 s & 11.612 s \\
\(A_7B_6C_2\) & FIgenerator & 0.070 s & 2.940 s & 101.180 s & 3,043.720 s \\
\(A_7B_6C_2\) & CMG & 0.033 s & 0.342 s & 4.640 s & 70.830 s \\
\end{longtable}
\endgroup

These results show that, at the commonly used degree-five and degree-six cutoffs, CMG completes offline generation for these small-molecule permutation groups in seconds to minutes.
Generation times for both types of program increase with degree and the number of element-color combinations.

\subsection{2 Generation time depends weakly on system size}\label{sec:size-independent-generation}

To examine the system-size dependence of generation time at fixed degree, we construct a sequence containing \(2^n\) water molecules and generate all global PIPs through degree five without component-count or treewidth pruning.
The numbers of oxygen atoms, hydrogen atoms, and total atoms are \(2^n\), \(2^{n+1}\), and \(3\times2^n\), respectively. These tests also use a single thread.

\begingroup
\small
\setlength{\tabcolsep}{4pt}
\begin{longtable}[]{@{}rrrrr@{}}
\toprule\noalign{}
\(n\) & Water molecules & Total atoms & Degree-five outputs & CMG generation time \\
\midrule\noalign{}
\endhead
\bottomrule\noalign{}
\endlastfoot
0 & 1 & 3 & 33 & 0.015 s \\
1 & 2 & 6 & 608 & 0.110 s \\
2 & 4 & 12 & 1,827 & 1.034 s \\
3 & 8 & 24 & 2,180 & 2.365 s \\
4 & 16 & 48 & 2,183 & 2.951 s \\
5 & 32 & 96 & 2,183 & 2.951 s \\
19 & 524,288 & 1,572,864 & 2,183 & 2.973 s \\
20 & 1,048,576 & 3,145,728 & 2,183 & 2.975 s \\
\end{longtable}
\endgroup

The output count reaches 2,183 at 16 water molecules and does not grow thereafter; generation time also stabilizes near 2.9 s.
At fixed degree five, each graph pattern requires only a bounded number of vertices of each color. Once the oxygen and hydrogen capacities exceed these bounds, color exhaustion removes no further patterns, and additional atoms do not create new CMG isomorphism classes.
For fixed element types and algebraic degree, the offline generation cost of CMG is therefore determined primarily by the graph-pattern space and depends only weakly on the number of atoms.

The following table lists the degree-five water-system cases completed by FIgenerator and the CMG times for the same systems. At \(n=4\), FIgenerator can no longer generate the corresponding PIPs within a reasonable time.
Generation time for explicit monomial-orbit expansions grows rapidly with system capacity, whereas CMG soon reaches a plateau by generating at the graph-pattern level.
(For \(n=0\), \texttt{0.000\ s} means that the runtime is below the resolution of \texttt{clock()} in that run, rather than that the computational cost is exactly zero.)
\begingroup
\small
\setlength{\tabcolsep}{4pt}
\begin{longtable}[]{@{}
  >{\raggedleft\arraybackslash}p{(\linewidth - 10\tabcolsep) * \real{0.1667}}
  >{\raggedleft\arraybackslash}p{(\linewidth - 10\tabcolsep) * \real{0.1667}}
  >{\raggedleft\arraybackslash}p{(\linewidth - 10\tabcolsep) * \real{0.1667}}
  >{\raggedleft\arraybackslash}p{(\linewidth - 10\tabcolsep) * \real{0.1667}}
  >{\raggedleft\arraybackslash}p{(\linewidth - 10\tabcolsep) * \real{0.1667}}
  >{\raggedleft\arraybackslash}p{(\linewidth - 10\tabcolsep) * \real{0.1667}}@{}}
\toprule\noalign{}
\begin{minipage}[b]{\linewidth}\raggedleft
\(n\)
\end{minipage} & \begin{minipage}[b]{\linewidth}\raggedleft
Water molecules
\end{minipage} & \begin{minipage}[b]{\linewidth}\raggedleft
Total atoms
\end{minipage} & \begin{minipage}[b]{\linewidth}\raggedleft
Degree-five outputs
\end{minipage} & \begin{minipage}[b]{\linewidth}\raggedleft
FIgenerator cumulative CPU time
\end{minipage} & \begin{minipage}[b]{\linewidth}\raggedleft
CMG generation time
\end{minipage} \\
\midrule\noalign{}
\endhead
\bottomrule\noalign{}
\endlastfoot
0 & 1 & 3 & 33 & 0.000 s & 0.015 s \\
1 & 2 & 6 & 608 & 0.020 s & 0.110 s \\
2 & 4 & 12 & 1,827 & 10.650 s & 1.034 s \\
3 & 8 & 24 & 2,180 & 55,024.990 s & 2.365 s \\
\end{longtable}
\endgroup

\subsection{3 Online evaluation on MD17}\label{sec:md17-evaluation}

\subsubsection{3.1 Complete PIPs versus conventional explicit expansions}\label{md17-full-pip}

We next compare online evaluation by CMG tensor-network descent with conventional explicit full-PIP formulas for the eight molecules in the original MD17 dataset \cite{Chmiela2017GDML}.
Both paths generate the complete PIP set through degree five and produce identical output counts for each system.
Tests use one logical CPU in the same environment as above, with all evaluations in fp64. CMG uses a JAX backend.
At these system sizes, conventional monomial expansions are scarcely compilable by machine-learning frameworks such as JAX, so the traditional formulas are evaluated using a handwritten C++ program.

The main test uses a batch size of 32. All reported times are divided by batch size and represent amortized time per configuration.

Measurements show that batch size has little effect on the MD17 comparison when evaluation uses a single CPU core.

CMG has two JAX backends that express the \(H_F\)-to-\(J_F\) descent differently: directly expanded calculations and sparse-matrix calculations in COO format.
The former is faster in online evaluation, while the latter compiles faster. Compilation is bottlenecked by the sparse linear transformation from \(H_F\) to \(J_F\).

Uracil uses the COO-JAX backend because the expanded form compiles too slowly. The other seven systems use directly expanded CMG calculations, producing the same complete PIP sets as the traditional program.

\begingroup
\small
\setlength{\tabcolsep}{4pt}
\begin{longtable}[]{@{}
  >{\raggedright\arraybackslash}p{(\linewidth - 10\tabcolsep) * \real{0.2000}}
  >{\raggedleft\arraybackslash}p{(\linewidth - 10\tabcolsep) * \real{0.1000}}
  >{\raggedleft\arraybackslash}p{(\linewidth - 10\tabcolsep) * \real{0.1400}}
  >{\raggedleft\arraybackslash}p{(\linewidth - 10\tabcolsep) * \real{0.1900}}
  >{\raggedleft\arraybackslash}p{(\linewidth - 10\tabcolsep) * \real{0.2000}}
  >{\raggedleft\arraybackslash}p{(\linewidth - 10\tabcolsep) * \real{0.1700}}@{}}
\toprule\noalign{}
\begin{minipage}[b]{\linewidth}\raggedright
System
\end{minipage} & \begin{minipage}[b]{\linewidth}\raggedleft
Atoms
\end{minipage} & \begin{minipage}[b]{\linewidth}\raggedleft
Full PIPs
\end{minipage} & \begin{minipage}[b]{\linewidth}\raggedleft
CMG time/config.
\end{minipage} & \begin{minipage}[b]{\linewidth}\raggedleft
Legacy full-PIP time/config.
\end{minipage} & \begin{minipage}[b]{\linewidth}\raggedleft
Speedup
\end{minipage} \\
\midrule\noalign{}
\endhead
\bottomrule\noalign{}
\endlastfoot
aspirin & 21 & 18,467 & 554.505 us & 25.011765 s & 45,106.4 \\
benzene & 12 & 2,099 & 41.886 us & 49.548 ms & 1,182.9 \\
ethanol & 9 & 3,233 & 46.849 us & 2.313 ms & 49.4 \\
malon\-aldehyde & 9 & 8,605 & 153.648 us & 2.420 ms & 15.8 \\
naph\-thalene & 18 & 2,180 & 51.694 us & 3.405529 s & 65,878.1 \\
salicylic acid & 16 & 16,262 & 451.927 us & 885.535 ms & 1,959.5 \\
toluene & 15 & 2,167 & 47.341 us & 456.611 ms & 9,645.2 \\
uracil & 12 & 54,760 & 3.463658 ms & 50.552 ms & 14.6 \\
\end{longtable}
\endgroup

Theoretically, tensor descent scales as \(M^3\), whereas conventional explicit evaluation scales approximately as \(M^{10}\);
however, the measured speedups are smaller than the asymptotic \(M^7\) ratio implied by the leading powers.
One reason is the difference in the prefactors of the operation counts:
the prefactor for conventional expanded evaluation is approximately \(0.001\), whereas that of the present algorithm is close to \(10\).
In addition, system-size-independent overhead remains substantial for small systems.
For a system of approximately ten atoms, the \(M^3\) tensor-contraction component involves only about \(10^4\) floating-point operations;
when the PIP count is large, the sparse linear transformation \(H_F\rightarrow J_F\) can require more operations than these tensor contractions.
Consequently, the total cost for the MD17 systems is not yet fully dominated by the asymptotic \(M^3\) term.

\subsubsection{3.2 Connected CMG generators versus fundamental invariants}\label{md17-connected-fi}

The complete PIP set contains many more polynomials than may be needed in practice. Traditional algorithms usually replace it with fundamental invariants (FIs). As discussed below, connected CMG patterns can be viewed as their counterparts.
For these small molecules, color exhaustion allows some connected generators to be eliminated, so the connected-pattern count exceeds the FI count. We compare tensor-network contraction and descent for connected patterns with traditional FI evaluation on the same eight MD17 systems.
Tests again use one thread in the same environment. All JAX backends are compiled in directly expanded form.

\begingroup
\small
\setlength{\tabcolsep}{4pt}
\begin{longtable}[]{@{}
  >{\raggedright\arraybackslash}p{(\linewidth - 12\tabcolsep) * \real{0.200}}
  >{\raggedleft\arraybackslash}p{(\linewidth - 12\tabcolsep) * \real{0.110}}
  >{\raggedleft\arraybackslash}p{(\linewidth - 12\tabcolsep) * \real{0.090}}
  >{\raggedleft\arraybackslash}p{(\linewidth - 12\tabcolsep) * \real{0.090}}
  >{\raggedleft\arraybackslash}p{(\linewidth - 12\tabcolsep) * \real{0.190}}
  >{\raggedleft\arraybackslash}p{(\linewidth - 12\tabcolsep) * \real{0.180}}
  >{\raggedleft\arraybackslash}p{(\linewidth - 12\tabcolsep) * \real{0.140}}@{}}
\toprule\noalign{}
System & Conn. graphs & FIs & \shortstack[r]{Conn./\\FI} & Connected P time/config. & Legacy FI time/config. & Descriptive speedup \\
\midrule\noalign{}
\endhead
\bottomrule\noalign{}
\endlastfoot
aspirin & 5,002 & 4,428 & 1.130 & 225.466 us & 331.030 ms & 1,468.2 \\
benzene & 787 & 721 & 1.092 & 26.248 us & 6.450 ms & 245.7 \\
ethanol & 1,521 & 517 & 2.942 & 42.866 us & 266.927 us & 6.2 \\
malon\-aldehyde & 3,501 & 1,025 & 3.416 & 104.712 us & 224.896 us & 2.15 \\
naph\-thalene & 787 & 784 & 1.004 & 32.356 us & 123.723 ms & 3,823.8 \\
salicylic acid & 4,723 & 3,605 & 1.310 & 183.095 us & 50.873 ms & 277.9 \\
toluene & 787 & 774 & 1.017 & 29.955 us & 34.828 ms & 1,162.7 \\
uracil & 15,348 & 5,266 & 2.915 & 602.149 us & 2.871 ms & 4.77 \\
\end{longtable}
\endgroup

The CMG JAX backend remains substantially faster than traditional evaluation.
The speedups are smaller than in the preceding comparison because FIs are much shorter than full PIPs and greatly reduce both complexity and operation count. Tensor descent still scales as \(M^3\), whereas FI evaluation scales approximately as \(M^6\).
Connected-pattern evaluation is also faster than evaluation of the complete PIP set because retaining only connected patterns substantially reduces the work in the descent transform.

\subsection{4 System-size scan for water clusters}\label{sec:water-cluster-scan}

MD17 consists primarily of small molecules and does not yet reach the asymptotic system-size regime of fixed-degree-five tensor contractions.
We therefore also test online evaluation of the complete PIP set through degree five for systems containing 1--1,024 water molecules.
We record both the forward time for a scalar energy formed from the PIPs and the time for reverse-mode gradient evaluation. The input is a pair-variable matrix, and gradients are taken with respect to that matrix.
These timings reflect the cost of neural-network backpropagation through a PIP layer in practical applications.
Tests run on a single NVIDIA A100 GPU using JAX \texttt{float64}, with the median of seven runs reported.
Reverse propagation costs substantially more than forward evaluation, but remains affordable; their time ratio gradually stabilizes as the system grows.

\subsubsection{4.1 Batch size 1}\label{batch-size-1}

\begingroup
\small
\setlength{\tabcolsep}{4pt}
\begin{longtable}[]{@{}
  >{\raggedleft\arraybackslash}p{(\linewidth - 12\tabcolsep) * \real{0.1429}}
  >{\raggedleft\arraybackslash}p{(\linewidth - 12\tabcolsep) * \real{0.1429}}
  >{\raggedleft\arraybackslash}p{(\linewidth - 12\tabcolsep) * \real{0.1429}}
  >{\raggedleft\arraybackslash}p{(\linewidth - 12\tabcolsep) * \real{0.1429}}
  >{\raggedleft\arraybackslash}p{(\linewidth - 12\tabcolsep) * \real{0.1429}}
  >{\raggedleft\arraybackslash}p{(\linewidth - 12\tabcolsep) * \real{0.1429}}
  >{\raggedleft\arraybackslash}p{(\linewidth - 12\tabcolsep) * \real{0.1426}}@{}}
\toprule\noalign{}
\begin{minipage}[b]{\linewidth}\raggedleft
Water molecules
\end{minipage} & \begin{minipage}[b]{\linewidth}\raggedleft
Atoms
\end{minipage} & \begin{minipage}[b]{\linewidth}\raggedleft
Energy forward
\end{minipage} & \begin{minipage}[b]{\linewidth}\raggedleft
Forward config./s
\end{minipage} & \begin{minipage}[b]{\linewidth}\raggedleft
Gradient
\end{minipage} & \begin{minipage}[b]{\linewidth}\raggedleft
Gradient config./s
\end{minipage} & \begin{minipage}[b]{\linewidth}\raggedleft
\shortstack[r]{Gradient/\\forward}
\end{minipage} \\
\midrule\noalign{}
\endhead
\bottomrule\noalign{}
\endlastfoot
1 & 3 & 0.041 ms & 24,297.7 & 0.116 ms & 8,624.0 & 2.82 \\
2 & 6 & 0.801 ms & 1,248.1 & 0.994 ms & 1,006.3 & 1.24 \\
4 & 12 & 3.117 ms & 320.9 & 4.757 ms & 210.2 & 1.53 \\
8 & 24 & 4.274 ms & 234.0 & 6.975 ms & 143.4 & 1.63 \\
16 & 48 & 4.428 ms & 225.8 & 7.603 ms & 131.5 & 1.72 \\
32 & 96 & 4.435 ms & 225.5 & 10.157 ms & 98.5 & 2.29 \\
64 & 192 & 4.567 ms & 219.0 & 8.564 ms & 116.8 & 1.88 \\
128 & 384 & 4.601 ms & 217.3 & 8.545 ms & 117.0 & 1.86 \\
256 & 768 & 4.853 ms & 206.1 & 9.215 ms & 108.5 & 1.90 \\
512 & 1,536 & 6.711 ms & 149.0 & 14.168 ms & 70.6 & 2.11 \\
1,024 & 3,072 & 16.797 ms & 59.5 & 43.328 ms & 23.1 & 2.58 \\
\end{longtable}
\endgroup

Tensor contractions become dominant only beyond 256 molecules. This reflects the computational characteristics of the GPU, which does not perform well with small batches.

\subsubsection{4.2 Batched evaluation}\label{sec:batched-evaluation}

The following table reports batched GPU evaluation on the same hardware. Batch size is 64 except for the 1,024-water system, where it is reduced to 32 because of GPU memory limits.
Batching substantially amortizes fixed scheduling costs for small systems.

\begingroup
\small
\setlength{\tabcolsep}{4pt}
\begin{longtable}[]{@{}
  >{\raggedleft\arraybackslash}p{(\linewidth - 12\tabcolsep) * \real{0.1429}}
  >{\raggedleft\arraybackslash}p{(\linewidth - 12\tabcolsep) * \real{0.1429}}
  >{\raggedleft\arraybackslash}p{(\linewidth - 12\tabcolsep) * \real{0.1429}}
  >{\raggedleft\arraybackslash}p{(\linewidth - 12\tabcolsep) * \real{0.1429}}
  >{\raggedleft\arraybackslash}p{(\linewidth - 12\tabcolsep) * \real{0.1429}}
  >{\raggedleft\arraybackslash}p{(\linewidth - 12\tabcolsep) * \real{0.1429}}
  >{\raggedleft\arraybackslash}p{(\linewidth - 12\tabcolsep) * \real{0.1426}}@{}}
\toprule\noalign{}
\begin{minipage}[b]{\linewidth}\raggedleft
Water molecules
\end{minipage} & \begin{minipage}[b]{\linewidth}\raggedleft
batch size
\end{minipage} & \begin{minipage}[b]{\linewidth}\raggedleft
Energy forward
\end{minipage} & \begin{minipage}[b]{\linewidth}\raggedleft
Forward config./s
\end{minipage} & \begin{minipage}[b]{\linewidth}\raggedleft
Gradient
\end{minipage} & \begin{minipage}[b]{\linewidth}\raggedleft
Gradient config./s
\end{minipage} & \begin{minipage}[b]{\linewidth}\raggedleft
Peak GPU memory
\end{minipage} \\
\midrule\noalign{}
\endhead
\bottomrule\noalign{}
\endlastfoot
1 & 64 & 0.047 ms & 1,361,560 & 0.121 ms & 526,779 & \(<0.01\) GiB \\
2 & 64 & 0.440 ms & 145,396 & 1.534 ms & 41,718 & 0.14 GiB \\
4 & 64 & 1.291 ms & 49,578 & 13.843 ms & 4,623 & 0.59 GiB \\
8 & 64 & 1.758 ms & 36,401 & 19.377 ms & 3,303 & 0.70 GiB \\
16 & 64 & 1.927 ms & 33,218 & 19.211 ms & 3,331 & 0.79 GiB \\
32 & 64 & 2.323 ms & 27,545 & 19.786 ms & 3,235 & 0.86 GiB \\
64 & 64 & 3.312 ms & 19,322 & 21.522 ms & 2,974 & 1.08 GiB \\
128 & 64 & 6.658 ms & 9,613 & 30.620 ms & 2,090 & 1.46 GiB \\
256 & 64 & 22.506 ms & 2,844 & 73.137 ms & 875 & 4.41 GiB \\
512 & 64 & 118.853 ms & 538 & 346.817 ms & 185 & 18.31 GiB \\
1,024 & 32 & 383.558 ms & 83.4 & 1.114 s & 28.7 & 37.54 GiB \\
\end{longtable}
\endgroup

\subsubsection{4.3 Water-system performance on one logical CPU}\label{h2o-cpu-staged-reverse}

The same online evaluation and automatic differentiation are also tested on a single-core CPU backend.
For larger systems, doubling the atom count increases forward time by approximately a factor of seven, close to the predicted cubic scaling. The reverse-to-forward time ratio also gradually stabilizes.

\begingroup
\small
\setlength{\tabcolsep}{4pt}
\begin{longtable}[]{@{}
  >{\raggedleft\arraybackslash}p{(\linewidth - 12\tabcolsep) * \real{0.105}}
  >{\raggedleft\arraybackslash}p{(\linewidth - 12\tabcolsep) * \real{0.095}}
  >{\raggedleft\arraybackslash}p{(\linewidth - 12\tabcolsep) * \real{0.105}}
  >{\raggedleft\arraybackslash}p{(\linewidth - 12\tabcolsep) * \real{0.175}}
  >{\raggedleft\arraybackslash}p{(\linewidth - 12\tabcolsep) * \real{0.175}}
  >{\raggedleft\arraybackslash}p{(\linewidth - 12\tabcolsep) * \real{0.175}}
  >{\raggedleft\arraybackslash}p{(\linewidth - 12\tabcolsep) * \real{0.170}}@{}}
\toprule\noalign{}
Water mol. & Atoms & PIPs & BS1 forward/config. & BS1 reverse/config. & BS64 forward/config. & BS64 reverse/config. \\
\midrule\noalign{}
\endhead
\bottomrule\noalign{}
\endlastfoot
1 & 3 & 33 & 0.016019 ms & 0.031614 ms & 0.000513 ms & 0.001041 ms \\
2 & 6 & 608 & 0.063250 ms & 0.471319 ms & 0.015941 ms & 0.378941 ms \\
4 & 12 & 1,827 & 0.379666 ms & 3.418046 ms & 0.099455 ms & 3.011995 ms \\
8 & 24 & 2,180 & 0.589743 ms & 4.911698 ms & 0.176507 ms & 4.299896 ms \\
16 & 48 & 2,183 & 0.737441 ms & 6.003300 ms & 0.263465 ms & 5.113668 ms \\
32 & 96 & 2,183 & 1.451622 ms & 9.925987 ms & 0.948548 ms & 9.863802 ms \\
64 & 192 & 2,183 & 4.448371 ms & 29.462652 ms & 4.624545 ms & 32.108822 ms \\
128 & 384 & 2,183 & 22.774956 ms & 134.925561 ms & 25.978405 ms & 142.681873 ms \\
256 & 768 & 2,183 & 160.189500 ms & 802.006957 ms & 163.278035 ms & 790.563696 ms \\
512 & 1,536 & 2,183 & 1.166292 s & 5.196972 s & 1.118120 s & 4.888972 s \\
1,024 & 3,072 & 2,183 & 8.104482 s & 34.928551 s & 8.105704 s & 34.225964 s \\
\end{longtable}
\endgroup

\section{Discussion}\label{sec:discussion}

\subsection{Applying PIPs to larger systems}\label{sec:discussion-system-size}

When the mathematical representation of PIPs changes from explicit monomial expansions to colored multigraphs, PIP formula generation is freed from the constraints of system size. Online evaluation cost still grows substantially with system size, however, and becomes the true bottleneck limiting the reach of global PIPs without a distance cutoff.

Previously, around 20 atoms approached the soft limit of global PIP evaluation: calculation remained possible but became very slow, and explicit degree-five formulas could reach approximately 10 GB. At around 40 atoms, formula storage could approach the hard limit imposed by the memory of a typical single server. For tensor-network descent, 20 atoms is still a very small system; with many element types, fixed costs such as descent may even exceed the tensor contractions themselves. At around a hundred atoms, tensor-network descent has only just finished warming up.

PIPs are one of the principal approaches to constructing potential energy surfaces for reactions, dissociation, and long-range interactions in gas-phase chemical dynamics~\cite{Braams2009PIP,Qu2018PIPReview,Bowman2025PIPPerspective}. Within the range of gas-phase reactions and small clusters, tensor-network descent removes system size as a major obstacle to global PIP evaluation. Extension to larger condensed-phase systems will still require strategies commonly used in local potentials, such as distance cutoffs and atom-centered descriptions, to be incorporated into the PIP framework~\cite{Allen2021aPIP}.

\subsection{A size-independent bound and color explosion}\label{sec:discussion-pattern-count}

The CMG representation directly shows that, for a fixed number of element colors \(E\), degree cutoff \(d_{\max}\), and contraction-width cutoff, the number of colored multigraphs has a fixed upper bound independent of molecular system size. Adding atoms increases the sums and contraction-tensor dimensions for each pattern, but does not increase the number of possible patterns. This is the size-independent bound. In the absence of color exhaustion, the polynomial count equals this bound. There is a threshold degree \(d\) below which no color exhaustion occurs. At degree \(d\) and above, the number of exhausted patterns is nondecreasing with total edge multiplicity. We call this degree the exhaustion threshold.

One of the most fundamental difficulties in fitting potential energy surfaces is the curse of dimensionality. Classical techniques, such as interpolation and ordinary polynomial expansions, can describe three- or four-atom surfaces efficiently and accurately, but their computational costs quickly become unmanageable as the atom count increases. Modern fitting methods rely more heavily on machine learning to address this difficulty and have extended successfully to large molecules.
PIPs were originally linear bases for direct polynomial expansions, rather than neural-network descriptors. They can be viewed as the result of imposing permutation invariance on a polynomial expansion of the potential and equating the coefficients of many equivalent monomials. Each equivalence class of monomials with equal coefficients forms a PIP basis function. The size-independent bound shows that the number of degrees of freedom, or coefficients to be determined, in a degree-\(d\) expansion has an upper bound independent of system size. This provides a way to alleviate the curse of dimensionality independently of machine-learning techniques such as neural networks. It may be an important structural reason for the effectiveness of PIPs in potential energy surfaces of small and medium-sized molecules~\cite{Braams2009PIP,Xie2010MSA,Jiang2013PIPNN,Moberg2021PIPUnrestricted,Hao2025PIPFIAlgorithms}.

At fixed degree \(d\), however, the basis count grows rapidly with the number of elements \(E\), roughly as \(E^d\) to \(E^{2d}\). We call this color explosion. In practice, systems with more element types can still achieve good results using a similar number of polynomials at lower total degree, although the reason remains unclear. Color explosion prevents a compact universal PIP potential from covering the entire periodic table, a capability available in many current universal potentials.

\subsection{A new perspective on fundamental invariants}\label{sec:discussion-fundamental-invariants}

Fundamental invariants (FIs) are often used directly as neural-network inputs, replacing the complete linear PIP basis with a smaller generating set~\cite{Shao2016FINN,Chen2020FINN2,Fu2023FIReview}. FI descriptors are more compact and efficient, but mathematically they no longer span the potential by linear combinations alone; polynomial combinations are required. FIs are therefore usually combined with neural networks. Intuitively, FIs generally offer better accuracy at a fixed descriptor count, whereas PIPs offer better accuracy at a fixed degree cutoff; FIs have a substantial inference-speed advantage in almost all cases. Beyond six atoms, the cost of traditional online polynomial-descriptor evaluation often begins to exceed that of the associated neural network, and its share of the total cost grows with system size.

In practical calculations, complete FI generating sets can be constructed only for small systems; somewhat larger systems require a degree cutoff. The truncated set generally no longer generates the entire invariant polynomial ring. It guarantees generation only of all PIPs through degree \(d\), while many higher-degree PIPs cannot be obtained from the retained generators. The graph representation reveals the rapid growth of complete FI sets with system size: for a system of 1,000 water molecules, the number of generators far exceeds the number of atoms in the observable universe. Degree truncation in large systems therefore selects only a small fraction of the lowest-degree generators from a set too vast to grasp intuitively. Successful applications also show that effective FI-based potential fitting does not require the chosen generators to generate the entire invariant polynomial ring.

Since practical fitting does not require a complete generating set, generator selection can also account for expressive power within the truncation and computational cost. Just as a linear space has many possible bases, an invariant polynomial algebra can have different irredundant generating sets~\cite{Thiery2000GraphInvariants}. The colored-multigraph language makes it straightforward to see that disconnected PIPs can be generated by polynomial combinations of connected PIPs of no greater degree. In the low-degree range with sufficient color capacity, connected graph orbit sums form a natural irredundant set of algebra generators. Finite-capacity color exhaustion introduces additional relations that allow some generators to be eliminated~\cite{Thiery2000GraphInvariants}. In other words, below the exhaustion threshold, connected graphs constitute one choice of FIs. For large systems, connected graphs are nearly equivalent to such a choice; for small systems, color exhaustion allows some connected graphs to be removed from the generating set.

Connected graphs in the present method need not serve as independent descriptors: they can also be used to construct disconnected PIP basis functions at low cost. Under \hyperref[sec:sloped-truncation]{sloped truncation}, FIs or connected graphs serve both as descriptors and as computational intermediates, seeking to combine FI inference speed with the linear expressive power of selected PIPs. Relative to an FI basis, this increases the input dimension and hence the number of network parameters. The Levenberg--Marquardt (LM) algorithm is currently the main optimizer used for PIP-NN training; increasing the parameter count places a substantial burden on training, while its impact on inference is more modest. Connected graphs thus serve both as compact descriptors and as shared intermediates, providing an adjustable tradeoff among evaluation cost, the linear PIP function space, and model parameter count.

\begingroup
\widowpenalty=10000
\clubpenalty=10000
\section{Conclusion}\label{sec:conclusion}

We have established a correspondence between PIP monomial orbits and colored multigraph isomorphism classes, 
and developed graph-pattern generation and tensor-network descent on this basis. For fixed element types and degree cutoff, 
offline generation is determined by a finite graph-pattern library and is no longer constrained by the number of atoms. 
Online evaluation recovers the complete PIP basis through shared tensor contractions and quotient-graph descent, 
with the principal system-size complexity controlled by graph contraction width. 
For the commonly used complete basis through degree five, evaluation falls from the tenth-power scaling of conventional explicit index enumeration to cubic scaling.

Numerical experiments confirm the computational advantages of this representation. Degree-five graph patterns for capacities of millions of atoms are generated in only a few seconds. Complete PIP evaluation on MD17 is accelerated by one to nearly five orders of magnitude, and forward and input-gradient calculations for water systems extend to more than three thousand atoms. These results show that enormous explicit orbit expansions are not a necessary price for retaining the complete PIP basis.

The CMG representation also integrates PIPs naturally with modern tensor computation and automatic differentiation, and provides a unified language for global, atom-centered, bond-centered, and cluster-centered descriptions. Graph patterns have consistent feature semantics across systems, laying a foundation for transferable, trainable PIP potential models.
\par
\endgroup

\section*{Acknowledgment}
\begin{sloppypar}
This work was supported by the Strategic Priority Research Program of the Chinese Academy of Sciences (XDB0970203), the National Natural Science Foundation of China (22525304 and 22288201),  the Innovation Program for Quantum Science and Technology (2021ZD0303305), and the Dalian Innovation Support Program (2021RD05).
\end{sloppypar}

\bibliographystyle{unsrtnat}
\bibliography{references}

\end{document}